\documentclass[a4paper]{eccomas_2022}
\usepackage{graphicx}
\usepackage{amsmath}
\usepackage{amsfonts}
\usepackage{amssymb}
\usepackage{float}
\usepackage{subfig}
\usepackage{placeins} 

\title{LARGE-EDDY SIMULATIONS OF TURBULENT COMPRESSIBLE SUPERSONIC JET FLOWS USING DISCONTINUOUS GALERKIN METHODS}

\author{Diego F. Abreu$^{1,\dagger,*}$, Carlos Junqueira-Junior$^{2}$, Eron T. V. Dauricio$^{1,\ddagger}$ and João Luiz F. Azevedo$^{3}$}

\heading{Diego F. Abreu, Carlos Junqueira-Junior, Eron T. V. 
Dauricio and João Luiz F. Azevedo}

\address{$^{1}$ Instituto Tecnológico de Aeronáutica, 12228--900, São José dos Campos, SP, Brazil, $^{\dagger}$mecabreu@yahoo.com.br , $^{\ddagger}$eron.tiago90@gmail.com
\and
$^{2}$ Arts et Métiers Institute of Technology, DynFluid, CNAM, HESAM University, 151 Boulevard de l'Hôpital, 75013, Paris, France, junior.junqueira@ensam.eu
\and
$^{3}$ Instituto de Aeronáutica e Espaço, 12228--904, São José dos Campos, SP, Brazil, joaoluiz.azevedo@gmail.com}


\abstract{In this work, a discontinuous Galerkin scheme is employed to perform LES simulations of supersonic jet flows. A total of four simulations are performed with different meshes and order of accuracy. The number of degrees of freedom from the simulations varies from $50 \times 10^6$ to $400 \times 10^6$. The results indicate that by increasing the resolution of simulation, in general, the results got closer to experimental data. The jet lipline is the only region in which this behavior is not observed. It investigated a procedure of using lower-order simulations to initialize high-order simulations and consequently reduce the total time of the simulation using high-order schemes. This strategy is successful and allows to perform high-order simulations with only $5 \%$ more computational effort than a complete second-order simulation.}

\begin{document}

\section{INTRODUCTION}

The Reynolds-Averaged Navier-Stokes (RANS) formulation has difficulty representing some types of fluid motions predominantly governed by free shear flows or wall-bounded flows with separated boundary layers. This difficulty is related to constructive assumptions of the formulation, characterized by the modeling of all turbulent quantities. The recent progress of computational power is enabling the employment of large-eddy simulations (LES) to simulate the problems that RANS formulation fails to model important aspects of the flow. Recent studies show the capability of LES simulations for reproducing free shear layer \cite{Bres2019, Kumar2017} and detached flows \cite{Ghate2021, Masoudi2021}. Another advantage of using LES is its capability to produce high-frequency unsteady information, which is necessary for aerodynamics, acoustics, loads, and heat transfer analyses.

The authors are interested in the simulation of supersonic jet flows for performing aerodynamic analyses of the shear layer regarding velocity and pressure fluctuations to improve the design of nozzles and adjacent structures. Different numerical options are employed to obtain the solution of LES formulation for jet flows. For example, low-order accuracy \cite{Junior2018} and high-order accuracy \cite{BogeyBailly2010, Debonis2017} finite difference schemes on structured meshes were employed to perform LES simulations of subsonic and supersonic jet flows. Low-order finite volume approach on unstructured meshes \cite{Mendezetal2012, Bres2017} is another option employed.

Due to the employment of structured meshes, the finite difference schemes have difficulty handling complex geometries. The finite volume schemes are applied to unstructured meshes, which make it easier to represent complex geometries, however, it is difficult to implement high-order discretizations with these schemes \cite{HesthavenWarburton2008}. In this context, the discontinuous Galerkin schemes are gaining relevance, because they are easily implemented with high-order accuracy discretizations and can be employed with unstructured meshes. Some work is already simulating jet flows with discontinuous Galerkin schemes \cite{Abreu2021, Corriganetal2018} or using similar strategies, for example, the Flux Reconstruction schemes \cite{ShenMiller2019}.

The discontinuous Galerkin schemes have multiple options for implementation. For example, one may choose to represent the solution by nodal or modal polynomials. It is possible to choose between different options of test functions that could better suits different types of elements, which are utilized to discretize the computational domain. One set of choices for the discontinuous Galerkin formulation is named discontinuous Galerkin spectral element method (DGSEM) \cite{Kopriva2010, Hindenlang2012}. The DGSEM, implemented in a numerical framework called FLEXI \cite{Krais2021}, was investigated for performing LES of a supersonic round jet flows with Mach number equal to $1.4$ and Reynolds number based on jet inlet diameter of $1.58 \times 10^6$ \cite{Abreu2021}.

The simulations were performed with two numerical meshes with $6.2 \times 10^6$ and $1.8 \times 10^6$ elements with second-order and third-order accurate discretizations, respectively. The two simulations were performed with nearly $50 \times 10^6$ degrees of freedom (DOF). They presented similar results, with the simulation performed with third-order accuracy requiring twice the time to perform the same simulation time as the second-order accurate simulation. When comparing the results to experimental data, excessive dissipation is observed, which led to shorter potential cores. The potential core of the jet is the length in the centerline of the jet where the velocity reaches $0.95$ of jet velocity. Other aspects of the flow, for example, the root mean square (RMS) of velocity fluctuations in the centerline and lipline of the jet, also presented some differences with experimental data. 

In this work, the results obtained using a new mesh are presented. The new mesh has a larger refinement and improved topology than the meshes utilized in previous work. The new mesh is simulated with second-order and third-order accurate discretizations. Discussions regarding the quality and improvement of the simulations are presented. A discussion of computational efficiency utilizing discontinuous Galerkin methods is also performed to develop guidelines for future works.

\section{NUMERICAL FORMULATION}
\subsection{GOVERNING EQUATIONS}
The work has an interest in the solution of the filtered Navier-Stokes equations. The filtering strategy is based on a spatial filtering process that separates the flow into a resolved part $\bar{( \cdot )}$ and a non-resolved part $( \cdot )'$. Implicit filter size is obtained from the mesh size. The filtered Navier-Stokes equations in conservative form can be written by
\begin{equation}
\frac{\partial \mathbf{\bar{Q}}}{\partial t} + \nabla \cdot \mathbf{F} ( \mathbf{\bar{Q}}, \nabla \mathbf{\bar{Q}})=0,
\label{eq.1}
\end{equation}
where $\mathbf{\bar{Q}}=[\bar{\rho}, \bar{\rho} \tilde{u}, \bar{\rho} \tilde{v}, \bar{\rho} \tilde{w}, \bar{\rho} \check{E}]^{T}$ is the vector of filtered conserved variables and $\mathbf{F}$ is the flux vector. The flux vector can be divided into the Euler fluxes and the viscous flux, $\mathbf{F}=\mathbf{F}^e-\mathbf{F}^v$. The fluxes with the filtered variables may be written as
\begin{equation}
\mathbf{F}_i^e= \left[ \begin{array}{c} 
                    \bar{\rho} \tilde{u}_i \\ \bar{\rho} \tilde{u} \tilde{u}_i + \delta_{1i} \bar{p}\\ \bar{\rho} \tilde{v} \tilde{u}_i + \delta_{2i}\bar{p} \\ \bar{\rho} \tilde{w} \tilde{u}_i + \delta_{3i}\bar{p} \\ (\bar{\rho} \check{E} + \bar{p}) \tilde{u}_i
                   \end{array} \right]  \hspace*{1.5 cm} 
\mathbf{F}_i^v= \left[ \begin{array}{c} 
                    0 \\ \tau_{1i}^{mod} \\ \tau_{2i}^{mod} \\ \tau_{3i}^{mod} \\ \tilde{u}_j \tau_{ij}^{mod} - q_i^{mod}
                   \end{array} \right] \hspace*{1.5 cm} \mbox{ , for } i = 1,2,3 ,                  
\end{equation}
where $\tilde{u}_i$ or $(\tilde{u}, \tilde{v}, \tilde{w})$ are the Favre averaged velocity components, $\bar{\rho}$ is the filtered density, $\bar{p}$ is the filtered pressure and $\bar{\rho} \check{E}$ is the filtered total energy per unit volume. The terms $\tau_{ij}^{mod}$ and $q_{i}^{mod}$ are the modified viscous stress tensor and heat flux vector, respectively, and $\delta_{ij}$ is the Kronecker delta. The filtered total energy per unit volume, according to the definition proposed by Vreman \cite{Vreman1995} in its "system I" approach, is given by
\begin{equation}
\bar{\rho} \check{E} = \frac{\bar{p}}{\gamma - 1} + \frac{1}{2}\bar{\rho}\tilde{u}_i\tilde{u}_i.
\end{equation}

The filtered pressure, Favre averaged temperature and filtered density are correlated using the ideal gas equation of state $\bar{p}= \bar{\rho} R \tilde{T}$, and $R$ is the gas constant, written as $R = c_p - c_v$. The properties $c_p$ and $c_v$ are the specific heat at constant pressure and volume, respectively. The modified viscous stress tensor may be written as
\begin{equation}
\tau_{ij}^{mod}=(\mu + \mu_{SGS}) \left(\frac{\partial \tilde{u}_i}{\partial x_j} + \frac{\partial \tilde{u}_j}{\partial x_i} \right) - \frac{2}{3} (\mu + \mu_{SGS}) \left(\frac{\partial \tilde{u}_k}{\partial x_k} \right) \delta_{ij} 
\end{equation}
where $\mu$ is the dynamic viscosity coefficient, calculated by Sutherland's Law, and $\mu_{SGS}$ is the SGS dynamic viscosity coefficient, which is provided by the subgrid-scale model. The strategy of modeling the subgrid-scale contribution as an additional dynamic viscosity coefficient is based on the Boussinesq hyphotesis. The modified heat flux vector, using the same modeling strategy, is given by
\begin{equation}
q_i^{mod}=-(k+k_{SGS})\frac{\partial \tilde{T}}{\partial x_i}
\end{equation}
where $k$ is the thermal conductivity coefficient of the fluid and $k_{SGS}$ is the SGS thermal conductivity coefficient given by
\begin{equation}
k_{SGS}=\frac{\mu_{SGS} c_p}{Pr_{SGS}}
\end{equation}
and $Pr_{SGS}$ is the SGS Prandtl number. The present work employs the static Smagorinsky model \cite{Smagorinsky1963} in order to calculate the subgrid-scale contribution.

\subsection{NODAL DISCONTINUOUS GALERKIN METHOD}
The nodal discontinuous Galerkin method used in this work is based on the modeling called discontinuous Galerkin spectral element method \cite{Kopriva2010, Hindenlang2012}. In this modeling strategy, the domain is divided into multiple hexahedral elements. This choice of elements permits the interpolating polynomial to be defined as a tensor product basis with degree $N$ in each space direction. This set of options leads to an algorithm that can be easily implemented and also produce a high level of computational efficiency due to reduced calculations.

In this method, the elements from the physical domain are mapped onto a reference unit cube elements $E=[-1,1]^3$. The equations, presented in (\ref{eq.1}) need also to be mapped to this new reference domain, leading to
\begin{equation}
J \frac{\partial \mathbf{\bar{Q}}}{\partial t} + \nabla_{\xi} \cdot \bar{\mathcal{F}} = 0,
\label{eq.2}
\end{equation}
where $\nabla_{\xi}$ is the divergence operator with respect to the reference element coordinates, $\mathbf{\xi}=(\xi^1,\xi^2,\xi^3)^T$, $J= \arrowvert \partial \mathbf{x} / \partial \mathbf{\xi} \arrowvert$ is the Jacobian of the coordinate transformation and $\bar{\mathcal{F}}$ is the contravariant flux vector.

The discontinuous Galerkin formulation is obtained multiplying (\ref{eq.2}) by the test function $\psi=\psi(\xi)$ and integrating over the reference element $E$
\begin{equation}
\int_E J \frac{\partial \mathbf{\bar{Q}}}{\partial t} \psi d \xi + \int_E \nabla_{\xi} \cdot \bar{\mathcal{F}} \psi d \xi = 0.
\label{eq.3}
\end{equation}
It is possible to obtain the weak form of the scheme by integrating by parts the second term in (\ref{eq.3})
\begin{equation}
\frac{\partial}{\partial t} \int_E J \mathbf{\bar{Q}} \psi d \xi + \int_{\partial E} (\bar{\mathcal{F}} \cdot \vec{N})^* \psi dS - \int_E \bar{\mathcal{F}} \cdot (\nabla_{\xi} \psi ) d \xi = 0,
\label{eq.4}
\end{equation}
where $\vec{N}$ is the unit normal vector of the reference element faces. Because the discontinuous Galerkin scheme allows discontinuities in the interfaces, the surface integral above is ill-defined. In this case, a numerical flux, $\bar{\mathcal{F}}^*$, is defined, and a Riemann solver is used to compute the value of this flux based on the discontinuous solutions given by the elements sharing the interface.

For the nodal form of the discontinuous Galerkin formulation, the solution in each element is approximated by a polynomial interpolation of the form
\begin{equation}
\mathbf{\bar{Q}}(\xi) \approx \sum_{p,q,r=0}^N \mathbf{\bar{Q}}_h(\xi_p^1,\xi_q^2,\xi_r^3,t)\phi_{pqr}(\xi),
\end{equation}
where $\mathbf{\bar{Q}}_h(\xi_p^1,\xi_q^2,\xi_r^3,t)$ is the value of the vector of conserved variables at each interpolation node in the reference element and $\phi_{pqr}(\xi)$ is the interpolating polynomial. For hexahedral elements, the interpolating polynomial is a tensor product basis with degree N in each space direction
\begin{equation}
\phi_{pqr}(\xi)=l_p(\xi^1)l_q(\xi^2)l_r(\xi^3), \hspace{10pt} l_p(\xi^1)= \prod_{\substack{i=0 \\ i \ne p}}^{N_p} \frac{\xi^1-\xi_i^1}{\xi_p^1-\xi_i^1}.
\end{equation}
The definitions presented are applicable to other two directions.

The numerical scheme used in the simulation additionally presents the split formulation \cite{Pirozzoli2011}, with the discrete form \cite{Gassner2016}, to enhance the stability of the simulation. The split formulation is employed for Euler fluxes only. The solution and the fluxes are interpolated and integrated at the nodes of a Gauss-Lobatto Legende quadrature, which presents the summation-by-parts property, that is necessary to employ the split formulation.

The Riemann solver used in the simulations is a Roe scheme with entropy fix \cite{Harten1983} to ensure that the second law of thermodynamics is respected, even with the split formulation. For the viscous flux, since the discontinuous Galerkin scheme is not suitable for discretizing the high order derivative operator, the lifting scheme of Bassi and Rebay \cite{BassiRebay1997} is used, which is also known for BR2. The time marching method chosen is a five-stage, fourth-order explicit Runge-Kutta scheme \cite{CarpenterKennedy1994}. The shock waves that appear in the simulation are stabilized using the finite-volume sub-cell shock-capturing method \cite{Sonntag2017}. The shock indicator of Jameson, Schmidt, and Turkel \cite{JST81} is utilized.

\section{EXPERIMENTAL CONFIGURATION}
The experimental work \cite{BridgesWernet2008} provides a good characterization of the flow properties for jet flows. Many configurations were analyzed. In this work, the interest is to simulate the fully expanded free jet flow configuration with a Mach number of $1.4$. In this configuration the jet flow has a static pressure in the nozzle exit section that equals the ambient static pressure with a supersonic velocity, for this reason, it is possible to avoid the use of nozzle wall geometries and also the shock waves are weaker when compared to other operating conditions.

The experimental apparatus for analyzed configuration is composed of a convergent-divergent nozzle designed with the method of characteristics \cite{BridgesWernet2008}. The nozzle exit diameter is $50.8$ mm. The Reynolds number based on nozzle exit diameter is approximately $1.58 \times 10^6$, which is large when compared to other jet experiments available in the literature.

The data acquisition in the tests applies Time-Resolved Particle Image Velocimetry (TRPIV) operated primarily with a $10$ kHz sample rate. The experiment uses two sets of cameras, one positioned to capture the flow along the nozzle centerline and the other positioned to capture the flow of the mixing layer along the nozzle lipline.

\section{NUMERICAL SETUP}
\subsection{GEOMETRY AND MESH CONFIGURATION}
\begin{figure}[htb!]
\centering
	\includegraphics[width=0.65\linewidth]{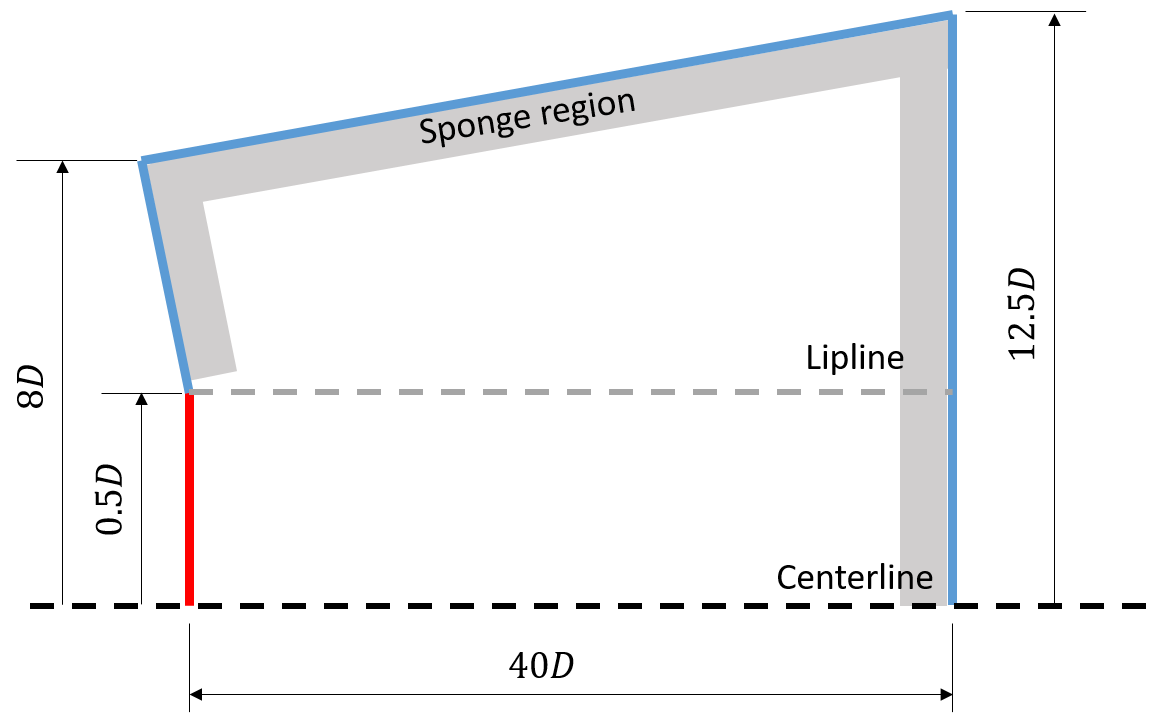}
    \caption{2-D schematic representation of the computational domain used on the
	jet flow simulations.}
    \label{fig.geo}
\end{figure}

The geometry used for the calculations in the work presents a divergent shape and axis length of $40D$, where $D$ is the jet inlet diameter and has external diameters of $16D$ and $25D$. Figure \ref{fig.geo} illustrates a 2-D representation of the computational domain indicating the inlet surface in red, the far-field region in blue, the lipline in gray, and the centerline in black. 

The computational grids used in the work are named M-1, M-2, and M-3. The M-1 and M-2 meshes are adaptations of the mesh utilized in previous work \cite{Junior2018} due to the different restrictions of each computational code. The M-3 mesh is generated with topological differences from M-1 and M-2 meshes. The M-3 mesh topology presents a high refinement level around the jet inlet boundary external diameter that transitions to a uniform distribution when moving forward in the longitudinal direction. In addition to the new topology, the M-3 mesh also presents a larger number of elements. The mesh generation uses a multiblock strategy since the FLEXI solver only handles hexahedral elements.

\begin{figure}[htb!]
\centering
\subfloat[M-2 mesh.]{
	\includegraphics[width=0.47\linewidth]{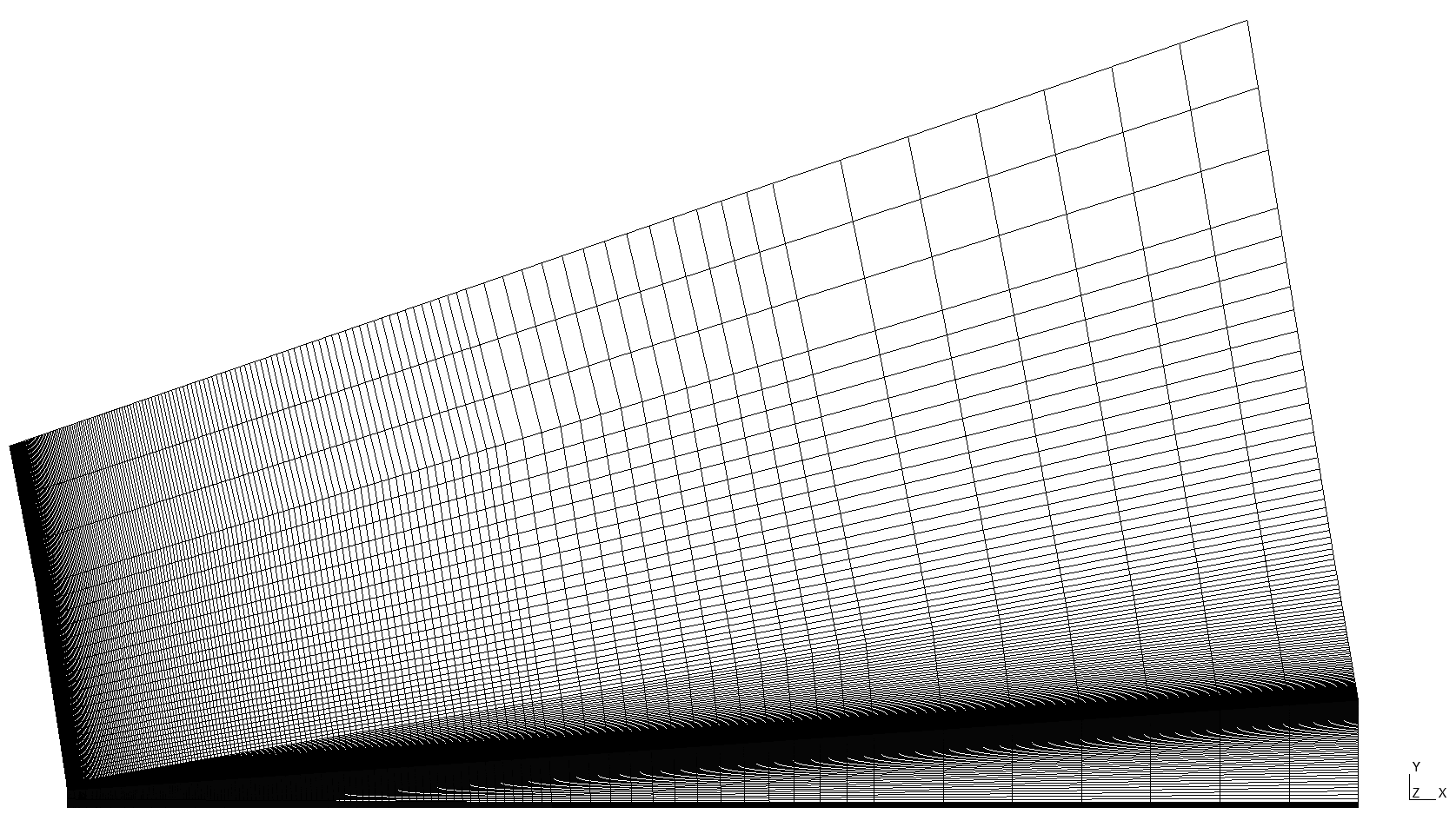}
	\label{fig.mes1}
	}
\subfloat[M-3 mesh.]{
	\includegraphics[width=0.47\linewidth]{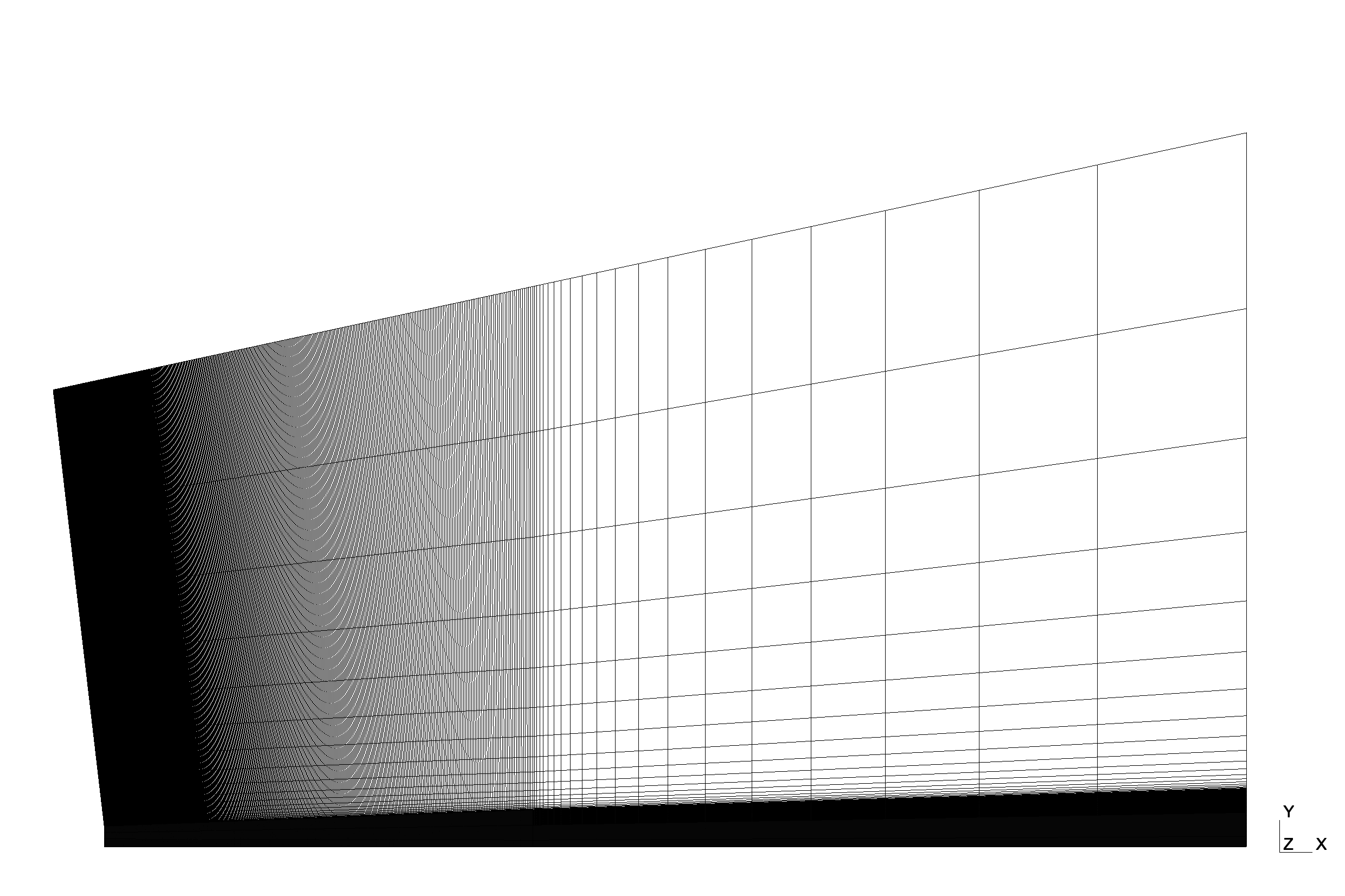}
	\label{fig.mesh2}
	}
\caption{Visualization of the half-plane longitudinal cut planes for the meshes used in the present work.}
\label{fig.mesh}
\end{figure}

Fig.\ \ref{fig.mesh} exhibits a cut plane of the M-2 and M-3 meshes. M-2 mesh is presented to illustrate the topological differences between the two strategies. M-1 mesh is not presented because it only differs in the number of elements from M-2 mesh. The M-1 and M-2 meshes have a total of $6.2 \times 10^6$ and $1.8 \times 10^6$ elements that are simulated with second and third-order accuracy, respectively, resulting in simulations with $50 \times 10^6$ DOF. The M-2 mesh has $15.4 \times 10^6$ elements and is simulated with second and third-order accuracy, resulting in approximately $120 \times 10^6$ and $410 \times 10^6$ DOF. All the meshes utilized in the work are generated with the GMSH \cite{Geuzaine2009} generator.

\subsection{BOUNDARY CONDITIONS}
different reference states to characterize the jet inflow, $(\cdot)_{jet}$, and the far-field, $(\cdot)_{ff}$. The inflow and the far-field surfaces are indicated in Fig.\ \ref{fig.geo} in red and blue, respectively. A weakly enforced solution of a Riemann problem with a Dirichlet condition is enforced at the boundaries. The flow is characterized as perfectly expanded and unheated, {\em i.e.} $p_{jet}/p_{ff}=T_{jet}/T_{ff}=1$, where $p$ stands for pressure and $T$ for temperature. The Mach number of the jet at the inlet is $M_{jet}=1.4$ and the Reynolds number based on the diameter of the nozzle is $Re_{jet} = 1.58 \times 10^6$. A small velocity component with $M_{ff}=0.01$ in the streamwise direction is imposed at the far-field to avoid numerical issues. A sponge zone \cite{Flad2014} is employed close to all far-field boundaries to dump any oscillation that could reach the boundaries.

\subsection{SIMULATION SETTINGS}
A total of 4 simulations are compared in this work. The development of the simulations utilized 3 different meshes with two orders of accuracy obtained by changing the degree of the polynomial representing the solution. The S-1 simulation utilizes the M-1 mesh with second-order accuracy. The S-2 simulation utilizes the M-2 mesh with third-order accuracy. The S-3 and S-4 simulations utilize the M-3 mesh with second and third-order accuracy, respectively. Table 1 summarizes the simulations performed and the total number of degrees of freedom in each of them.

\begin{table}[htb!]
\centering
\caption{Summary of simulations settings.}
\begin{tabular}{ c | c | c | c | c | c } \hline
Simulation & Meshes & Order of  & DOF/cell & Cells  & Total \# of DOF \\ 
 & & Accuracy & & ($10^{6}$) & ($10^{6}$) \\\hline
S-1 & M-1 & 2nd order & 8 & $6.2$ & $\approx 50$ \\
S-2 & M-2 & 3rd order & 27 & $1.8$ & $\approx 50$ \\
S-3 & M-3 & 2nd order & 8 & $15.4$ & $\approx 120$ \\ 
S-4 & M-3 & 3rd order & 27 & $15.4$ & $\approx 410$ \\ \hline
\end{tabular}
\label{tab.mesh}
\end{table}

\subsection{CALCULATION OF STATISTICAL PROPERTIES}
Two different approaches are taken to perform the 4 simulations. In the first approach, utilized for S-1, S-2, and S-3 simulations, the procedure involves three steps. The first one is to clean off the domain since the computation starts with a quiescent flow initial condition. The simulations run three flow-through times (FTT) to develop the jet flow. One FTT is the time required for one particle with the jet velocity to cross the computational domain. In the sequence, the simulations run an additional three FTT to produce a statistically steady condition. Then, in the last step, data are collected with a sample of approximately $250$ kHz for another FTT to obtain the statistical properties of the flow. In the second approach, utilized for the S-4 simulation, the solution obtained from the S-3 simulation is utilized as the initial condition. The simulation is performed for 0.5 FTT to clean the second-order accuracy solution and allow it to provide a third-order accuracy solution. Then 2 additional FTT are simulated to extract data for the analysis. The cost of S4 simulation is higher than other simulations and the authors had some difficulties to stabilize the simulation, which consumed some available computational resources, for this reason it was not possible to run 3 FTT to obtain the statistics.

\begin{figure}[htb!]
\centering
\includegraphics[width=0.8\linewidth]{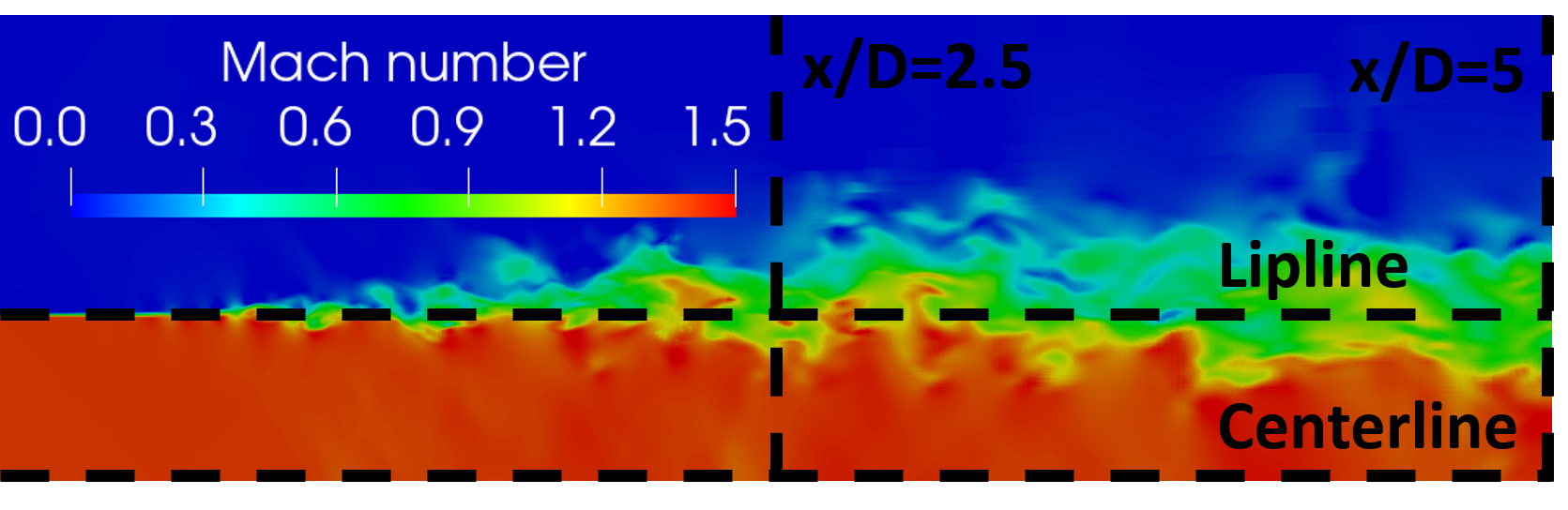}
\caption{Snapshot of the jet simulation with the two longitudinal lines and three crossflow lines 
along which data is extracted. Mach number contours are shown.}
\label{fig.jet_data_extract}
\end{figure}

The mean and the root mean square (RMS) fluctuations of properties of the flow are calculated along the centerline, lipline, and different domain surfaces in the streamwise direction. The centerline is defined as the line in the center of the geometry $y/D=0$, whereas the lipline is a surface parallel to the centerline and located at the nozzle diameter, $y/D=0.5$. The results from the lipline are an azimuthal mean from six equally spaced positions. The four surfaces in the streamwise positions are $x/D=2.5$, $x/D=5.0$, $x/D=10.0$, and $x/D=15.0$. Fig.\ \ref{fig.jet_data_extract} illustrates a snapshot of the jet flow with the lines and surfaces of data extraction. Mach number contours are presented in the figure.

\section{RESULTS}
\subsection{ANALYSIS OF NUMERICAL RESULTS}
The results from S1, S2, S3, and S4 simulations are presented in this section, which is divided into two parts to group different types of comparisons. In the first part, contours of mean longitudinal velocity, RMS longitudinal velocity fluctuation, and mean density are presented for each simulation. In the second part, the distribution of mean longitudinal velocity and RMS of longitudinal velocity fluctuation are presented along the jet centerline and lipline for the four simulations and compared to experimental data. In the final results, the mean longitudinal velocity, RMS of longitudinal velocity fluctuation, RMS of radial velocity fluctuation, and shear-stress tensor are presented in four spanwise lines for all the simulations and compared to experimental data.

In the first part, three main aspects can be analyzed from the different contours investigated and each contour is better to discuss one of the three aspects. The length of the potential core cannot be directly assessed from visual inspection, so the authors prefer to refer to the region of high velocity that can be easily inspected from the results of mean longitudinal velocity. The development of the shear layer can be visualized in all results, however, the one in which its intensity can be better visualized is in the results of RMS of longitudinal velocity fluctuation. The last aspect that can be assessed from the contours is the development of the series of shock and expansion waves in the early stages of the jet.

\begin{figure}[htb!]
\subfloat[S1 simulation.]{	
	\includegraphics[trim = 0mm 35mm 0mm 50mm, clip, width=0.48\linewidth]{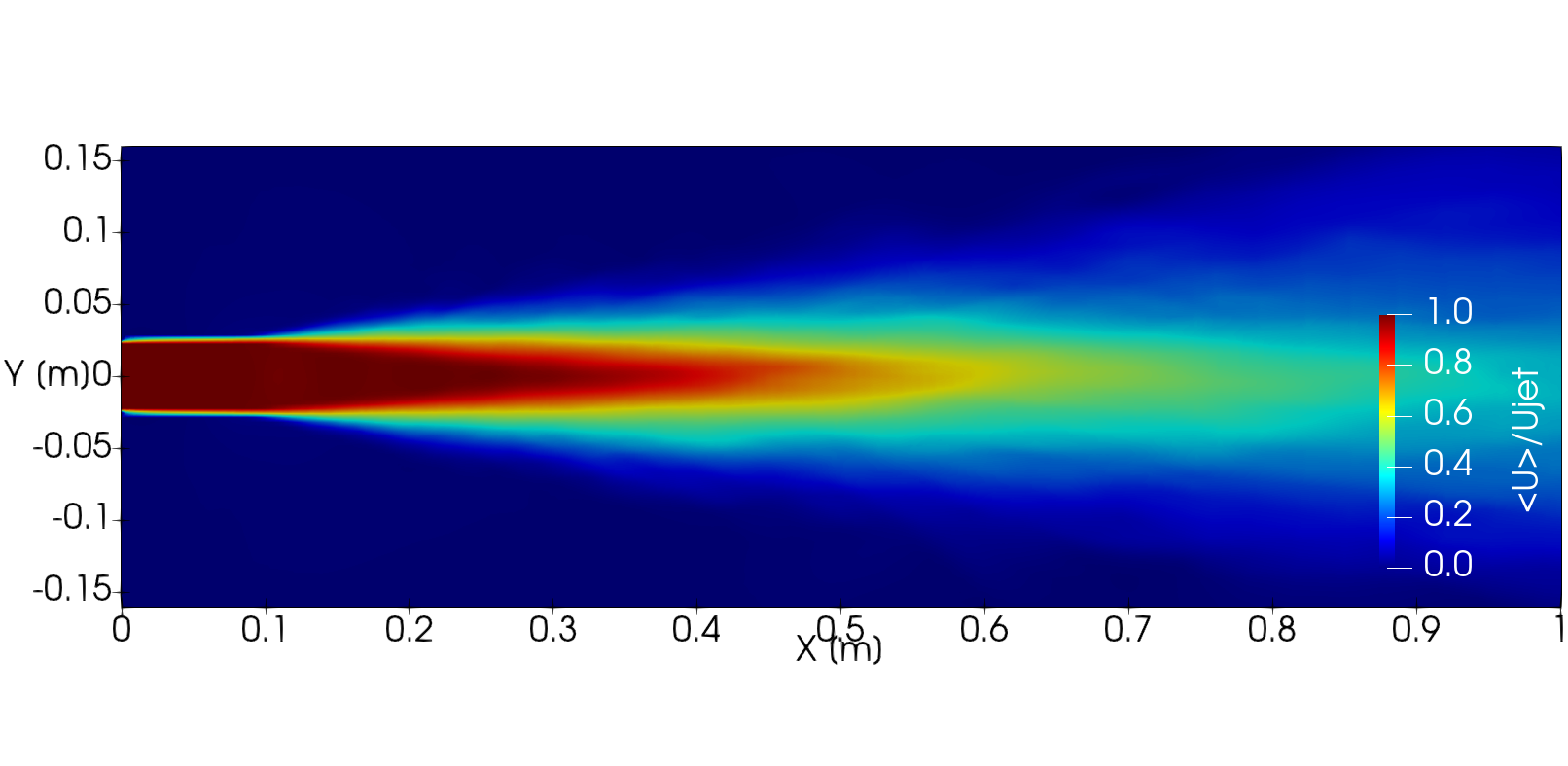}
	\label{fig.res_mvelxa}
	}
\subfloat[S2 simulation.]{
	\includegraphics[trim = 0mm 35mm 0mm 50mm, clip, width=0.48\linewidth]{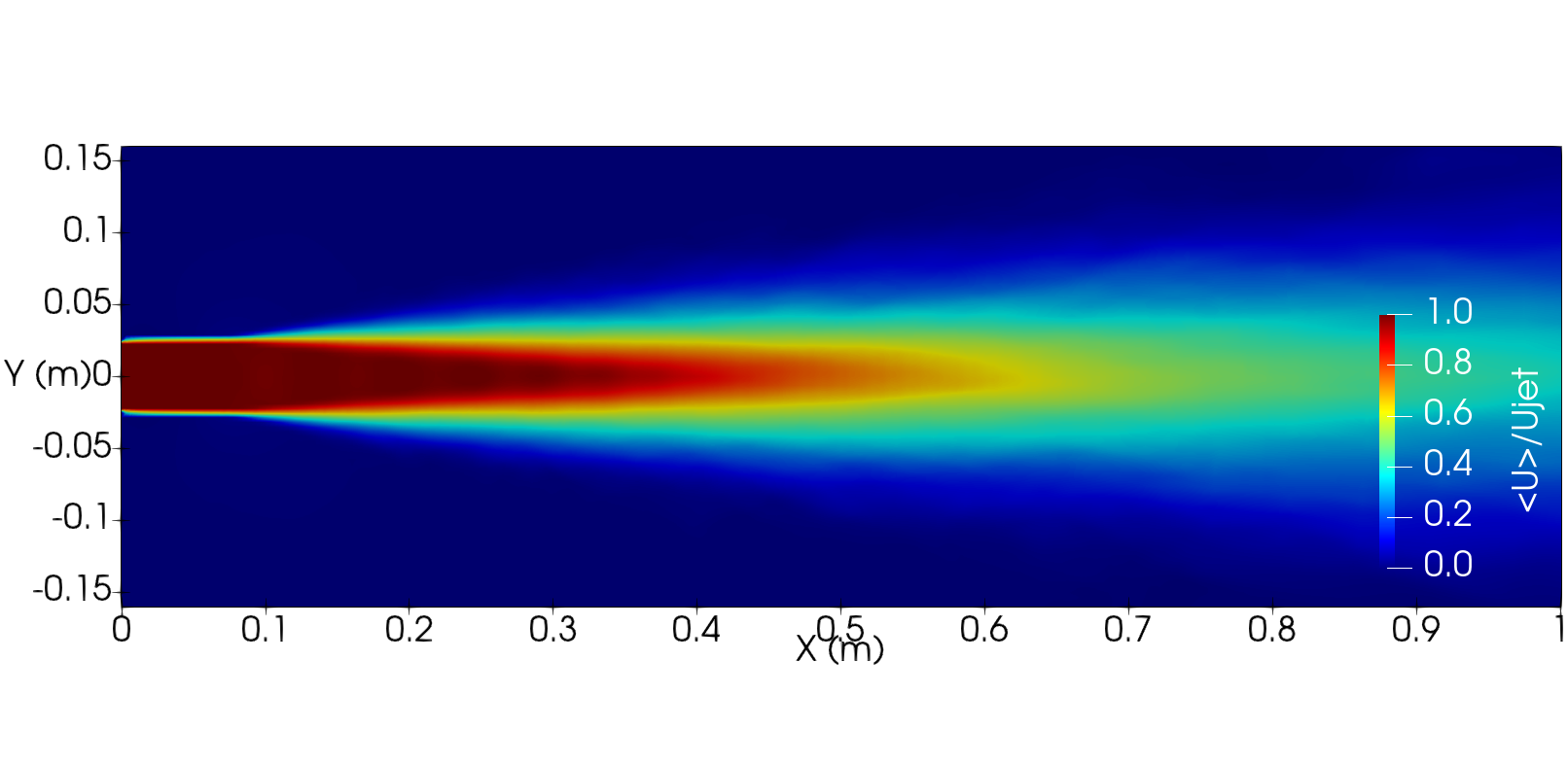}
	\label{fig.res_mvelxb}
	}
\\
\subfloat[S3 simulation.]{	
	\includegraphics[trim = 0mm 35mm 0mm 50mm, clip, width=0.48\linewidth]{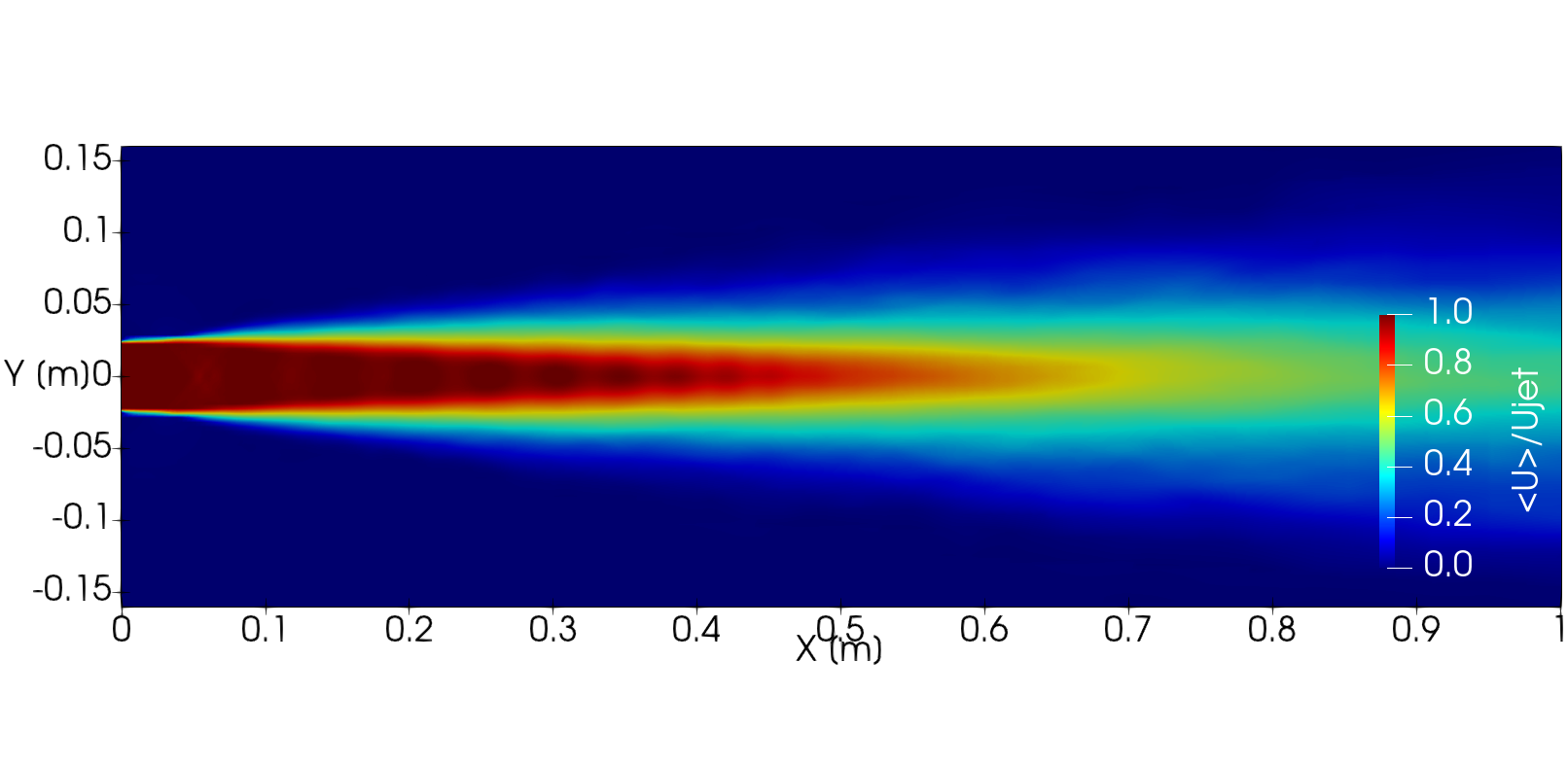}
	\label{fig.res_mvelxc}
	}
\subfloat[S4 simulation.]{
	\includegraphics[trim = 0mm 35mm 0mm 50mm, clip, width=0.48\linewidth]{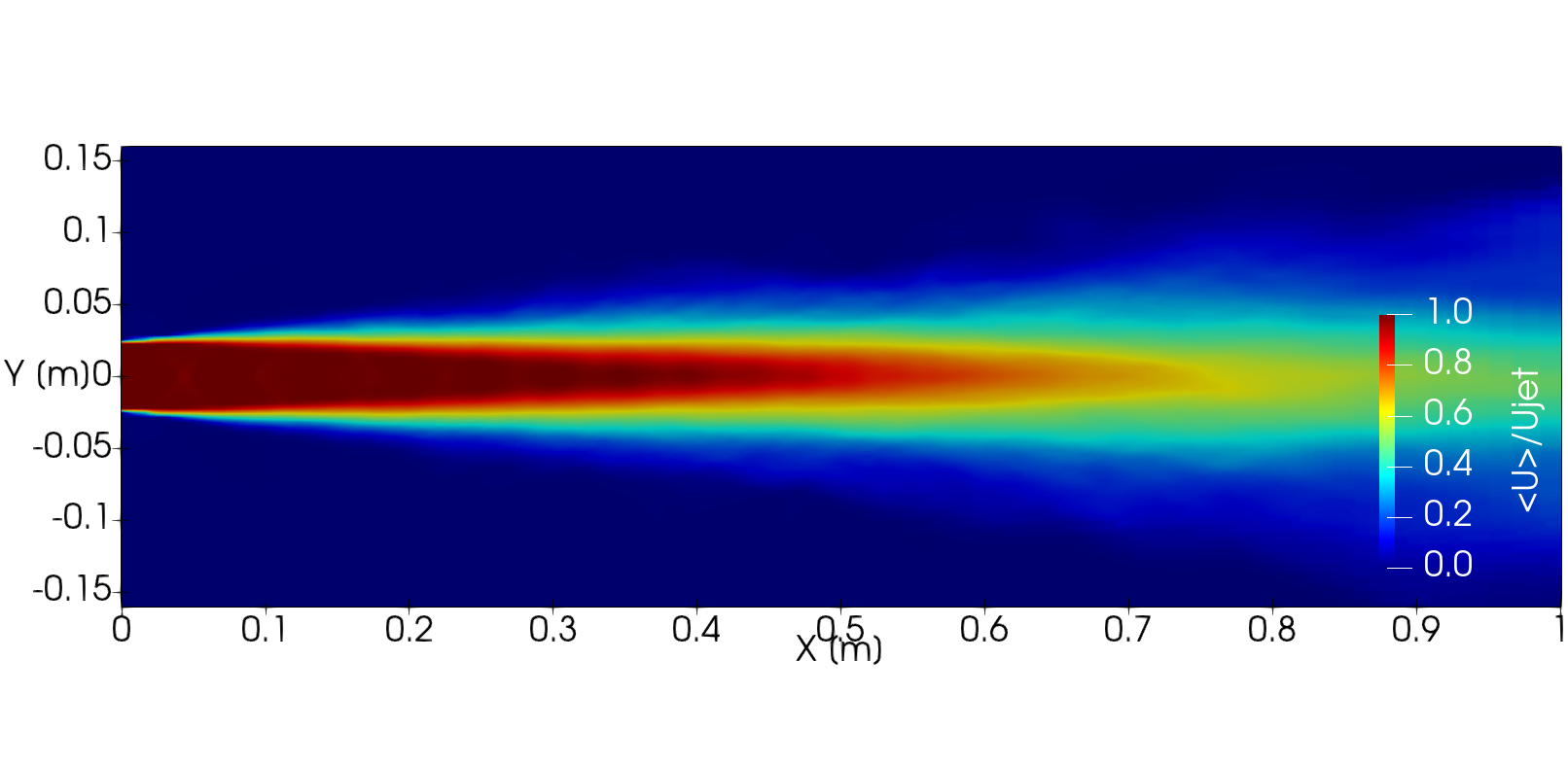}
	\label{fig.res_mvelxd}
	}
\caption{Contours of mean longitudinal velocity component on cutplane in $z/D=0$.}
\label{fig.res_mvelx}
\end{figure}

Figure \ref{fig.res_mvelx} presents the contours for the mean longitudinal velocity for all simulations. In Figs.\ \ref{fig.res_mvelxa} and \ref{fig.res_mvelxb} the contours of velocity are very similar, with the high velocity region in Fig.\ \ref{fig.res_mvelxb} being slightly longer. Analyzing the results in Fig.\ \ref{fig.res_mvelxc}, one can observe that the high-velocity region has increased significantly when compared to previous results. The improvement in the results obtained shows the importance of distributing elements where they are necessary. Finally, in Fig.\ \ref{fig.res_mvelxd}, the results from S4 simulation are presented. It is possible to observe that the high-velocity region is the longest among all the simulations, which is indicative that it was lacking resolution in previous simulations to adequately capture the development of the jet flow by adding too much dissipation.

\begin{figure}[htb!]
\subfloat[S1 simulation.]{	
	\includegraphics[trim = 0mm 35mm 0mm 50mm, clip, width=0.48\linewidth]{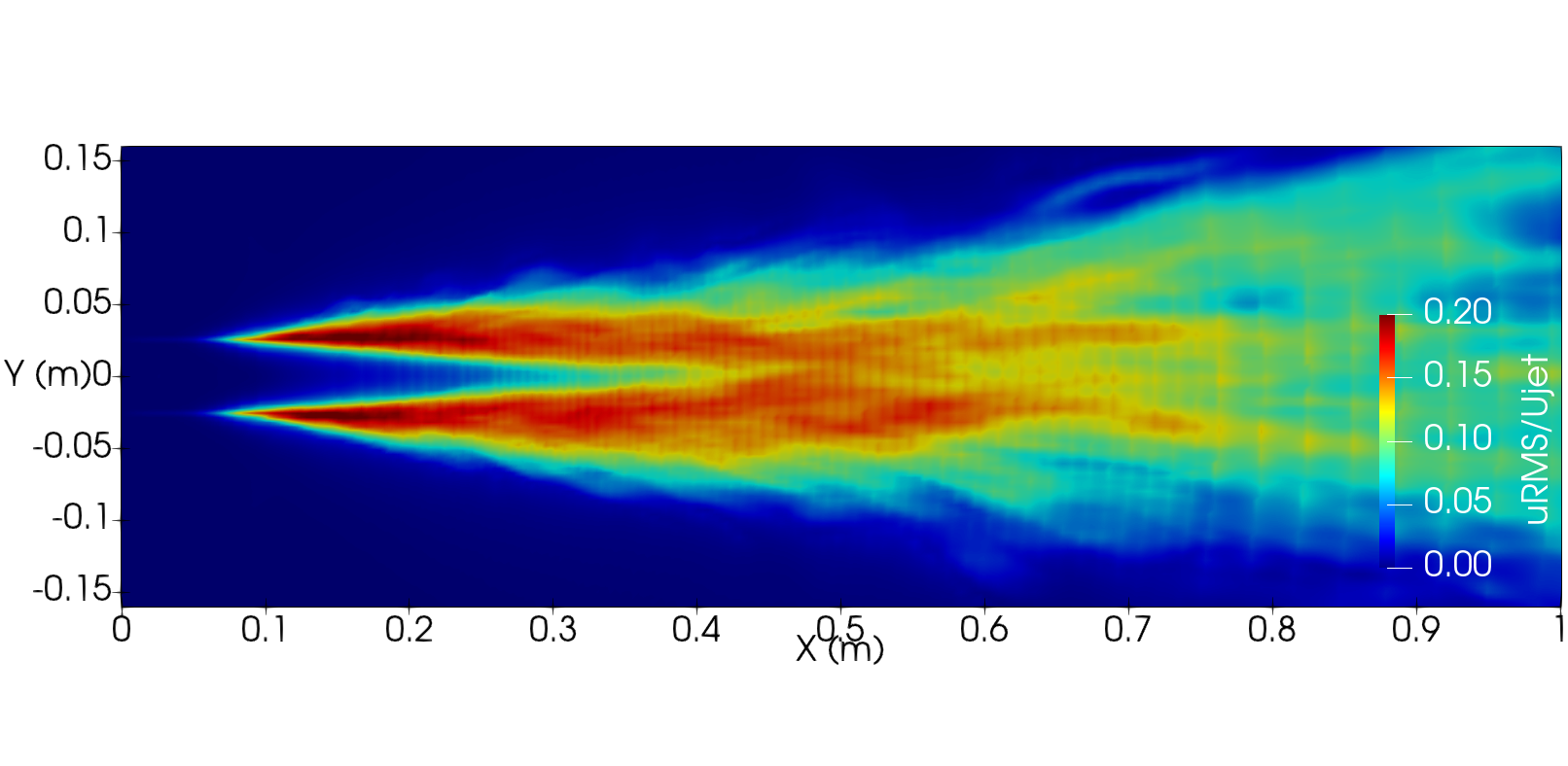}
	\label{fig.res_rvelxa}
	}
\subfloat[S2 simulation.]{
	\includegraphics[trim = 0mm 35mm 0mm 50mm, clip, width=0.48\linewidth]{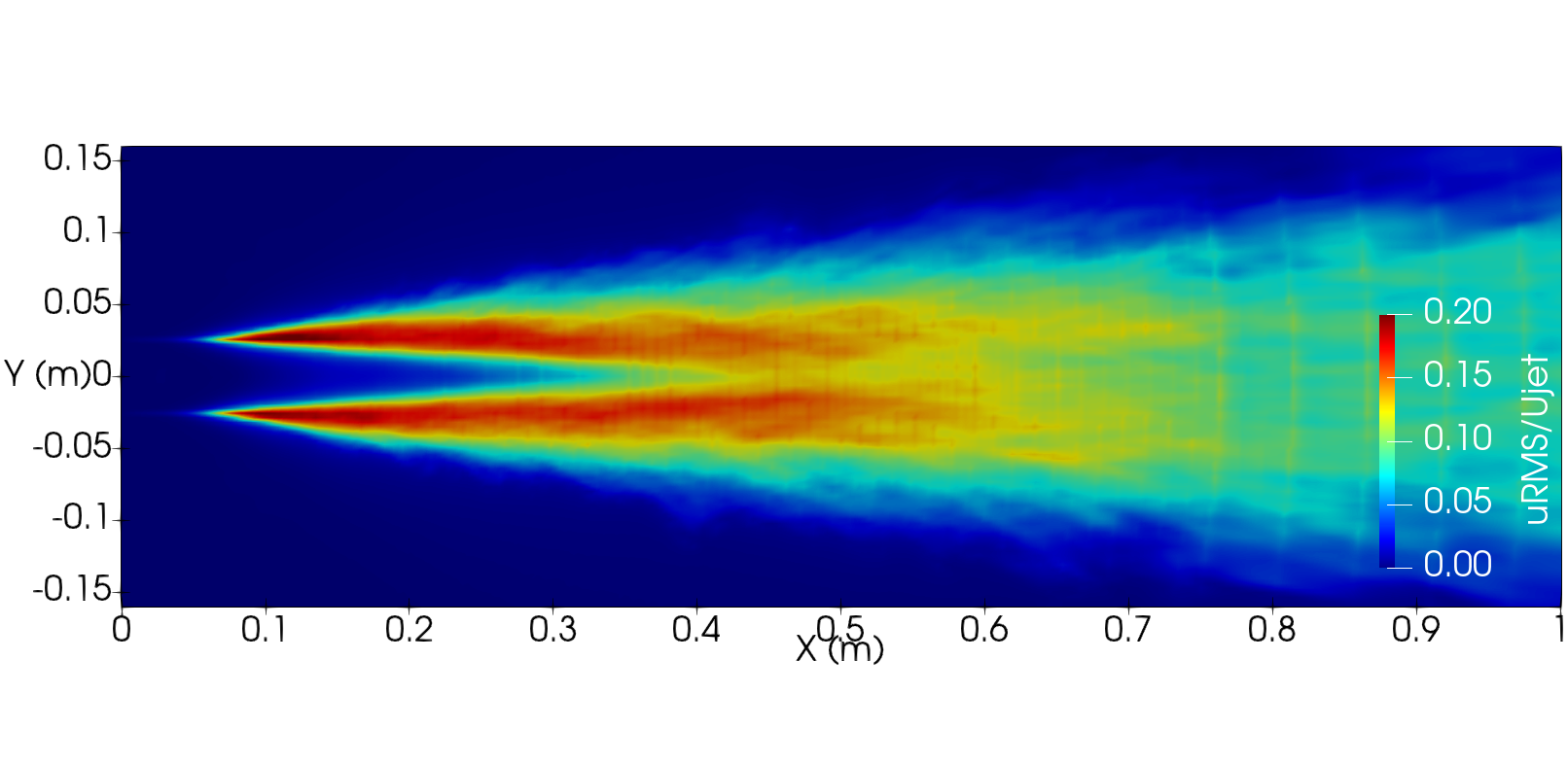}
	\label{fig.res_rvelxb}
	}
\\
\subfloat[S3 simulation.]{	
	\includegraphics[trim = 0mm 35mm 0mm 50mm, clip, width=0.48\linewidth]{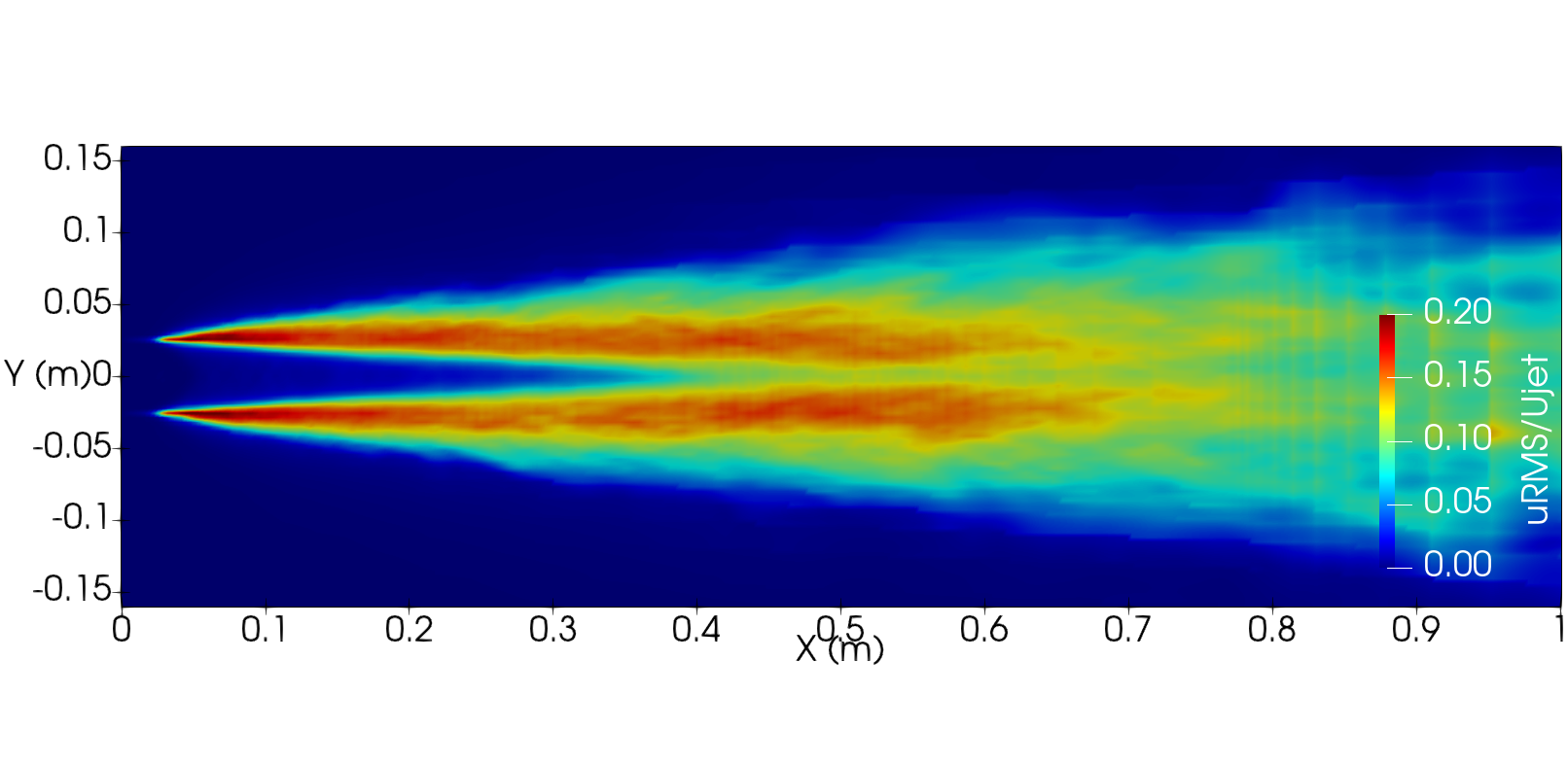}
	\label{fig.res_rvelxc}
	}
\subfloat[S4 simulation.]{
	\includegraphics[trim = 0mm 35mm 0mm 50mm, clip, width=0.48\linewidth]{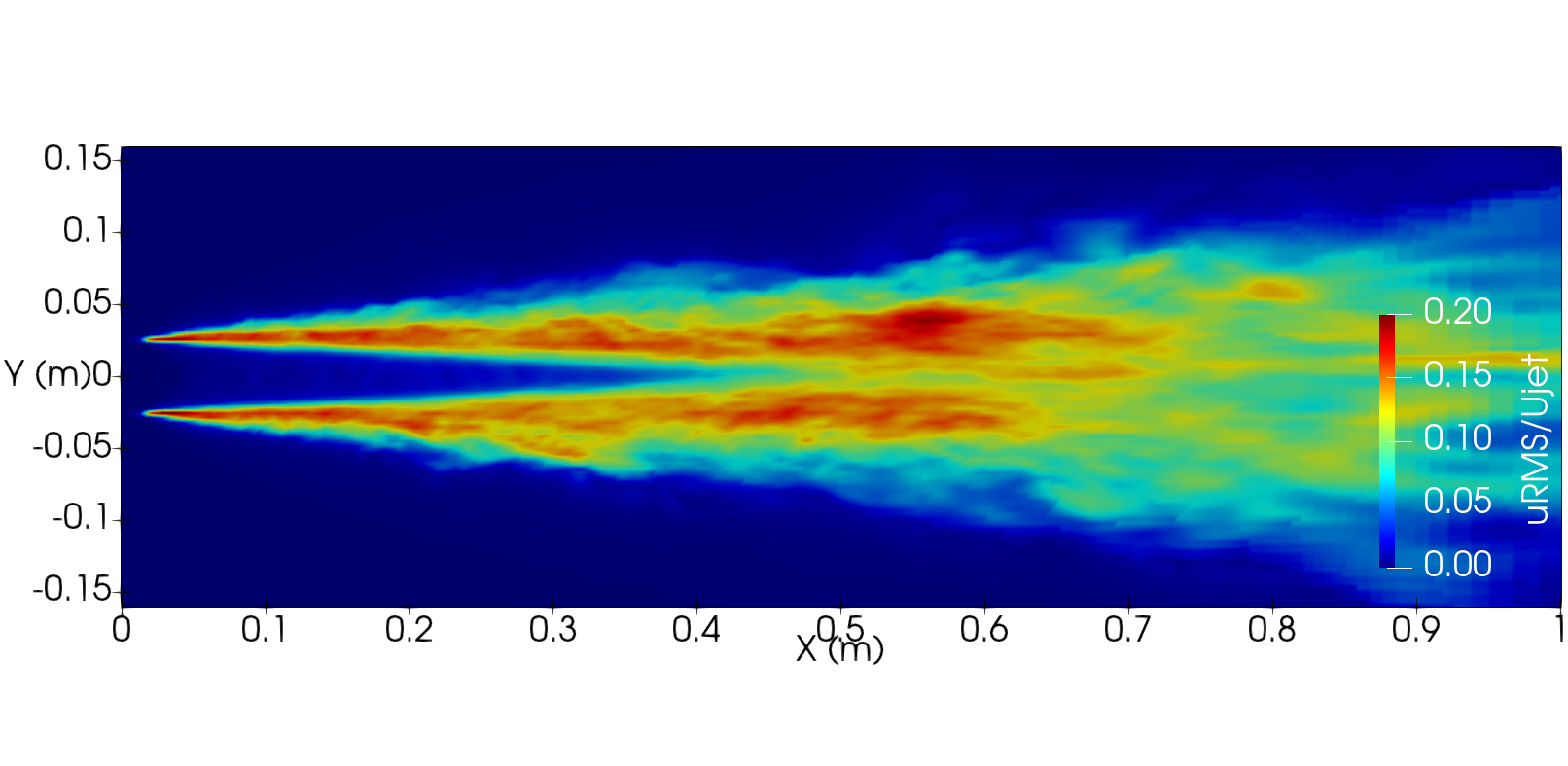}
	\label{fig.res_rvelxd}
	}
\caption{Contours of RMS of longitudinal velocity fluctuation on cutplane in $z/D=0$.}
\label{fig.res_rvelx}
\end{figure}

In Fig.\ \ref{fig.res_rvelx} the contours of RMS of longitudinal velocity fluctuation is presented. Once more, the results presented in Figs.\ \ref{fig.res_rvelxa} and \ref{fig.res_rvelxb} from S1 and S2 simulations are similar, with the shear-layer development starting approximately $1D$ far from the jet inlet section. Just after the initial of the shear layer development, one can observe that the peak of RMS fluctuation occurs, which can be related to the large difference between the velocities and possibly the transition of the shear layer from laminar to turbulent. The results presented in Fig.\ \ref{fig.res_rvelxc} from the S3 simulation have significant differences from the other two previously discussed. The development of the shear layer is starting closer to the jet inlet section with smaller peaks of RMS of velocity fluctuation and with a smaller spreading. One can visualize that the two mixing layers are crossing in the center of the jet farther in Fig.\ \ref{fig.res_rvelxc} than in Figs.\ \ref{fig.res_rvelxa} and \ref{fig.res_rvelxb} even presenting a sooner development. Analyzing the results in Fig.\ \ref{fig.res_rvelxd}, it can be observed that tendencies from the previous investigation continued to increase, which means that the beginning of the development of the shear layer got even closer to the jet inlet section and the crossing of the two mixing layers is happening farther from jet inlet section than the results from S3 simulation, Fig.\ \ref{fig.res_rvelxc}. S4 simulation is the one with smaller spreading and early development of the shear layer among all the simulations.

\begin{figure}[htb!]
\subfloat[S1 simulation.]{	
	\includegraphics[trim = 0mm 35mm 0mm 50mm, clip, width=0.48\linewidth]{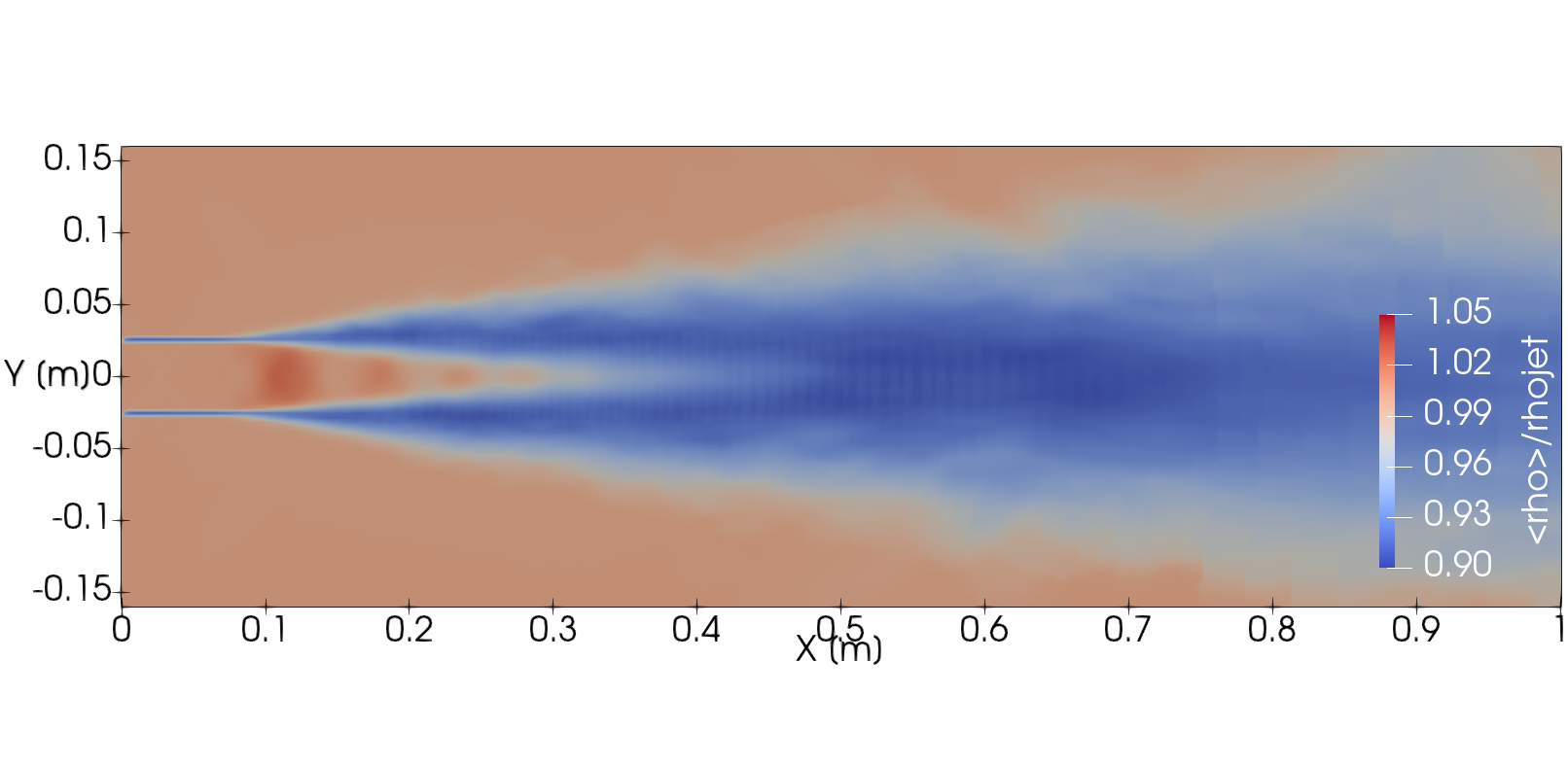}
	\label{fig.res_rhoa}
	}
\subfloat[S2 simulation.]{
	\includegraphics[trim = 0mm 35mm 0mm 50mm, clip, width=0.48\linewidth]{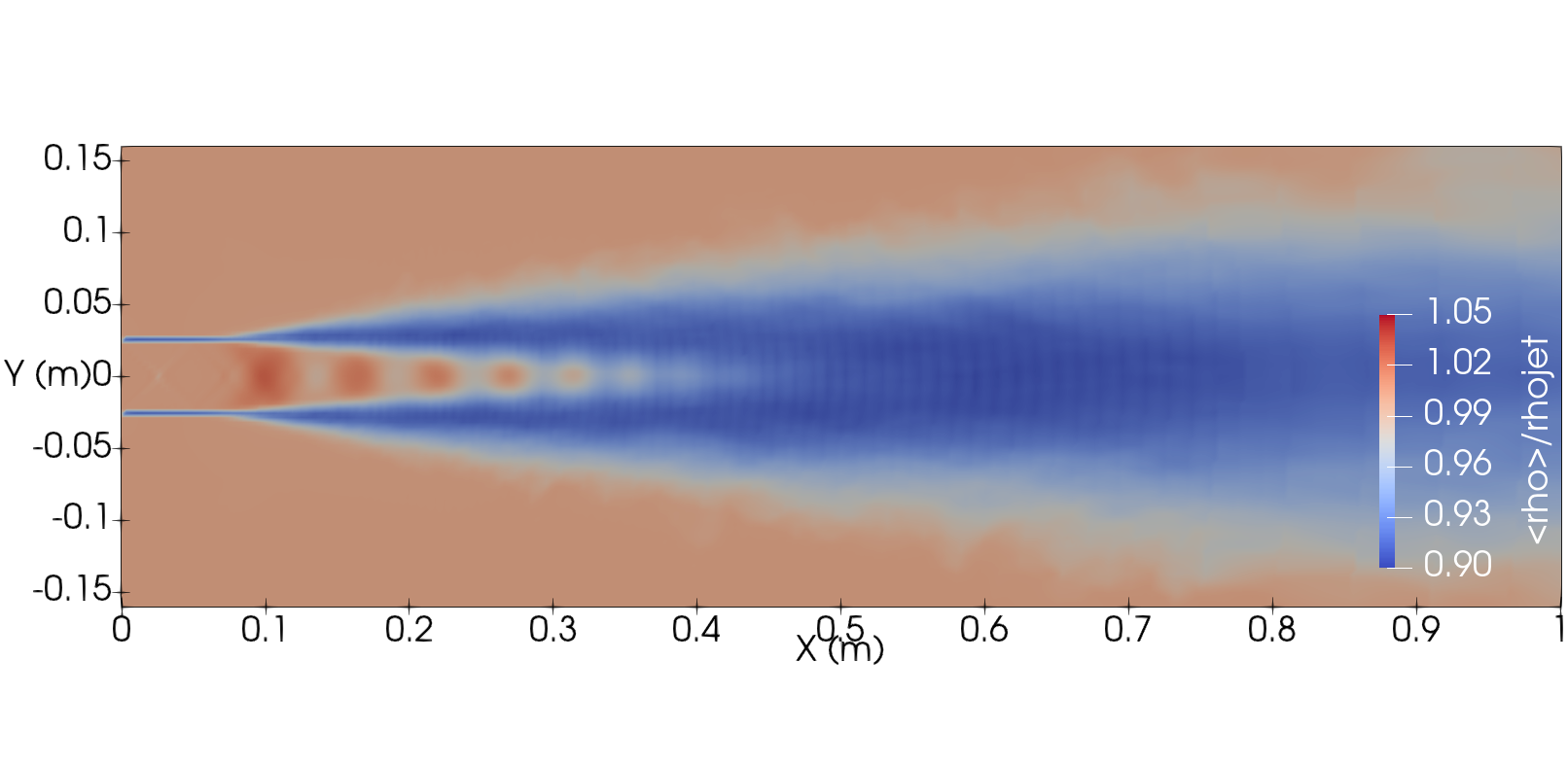}
	\label{fig.res_rhob}
	}
\\
\subfloat[S3 simulation.]{	
	\includegraphics[trim = 0mm 35mm 0mm 50mm, clip, width=0.48\linewidth]{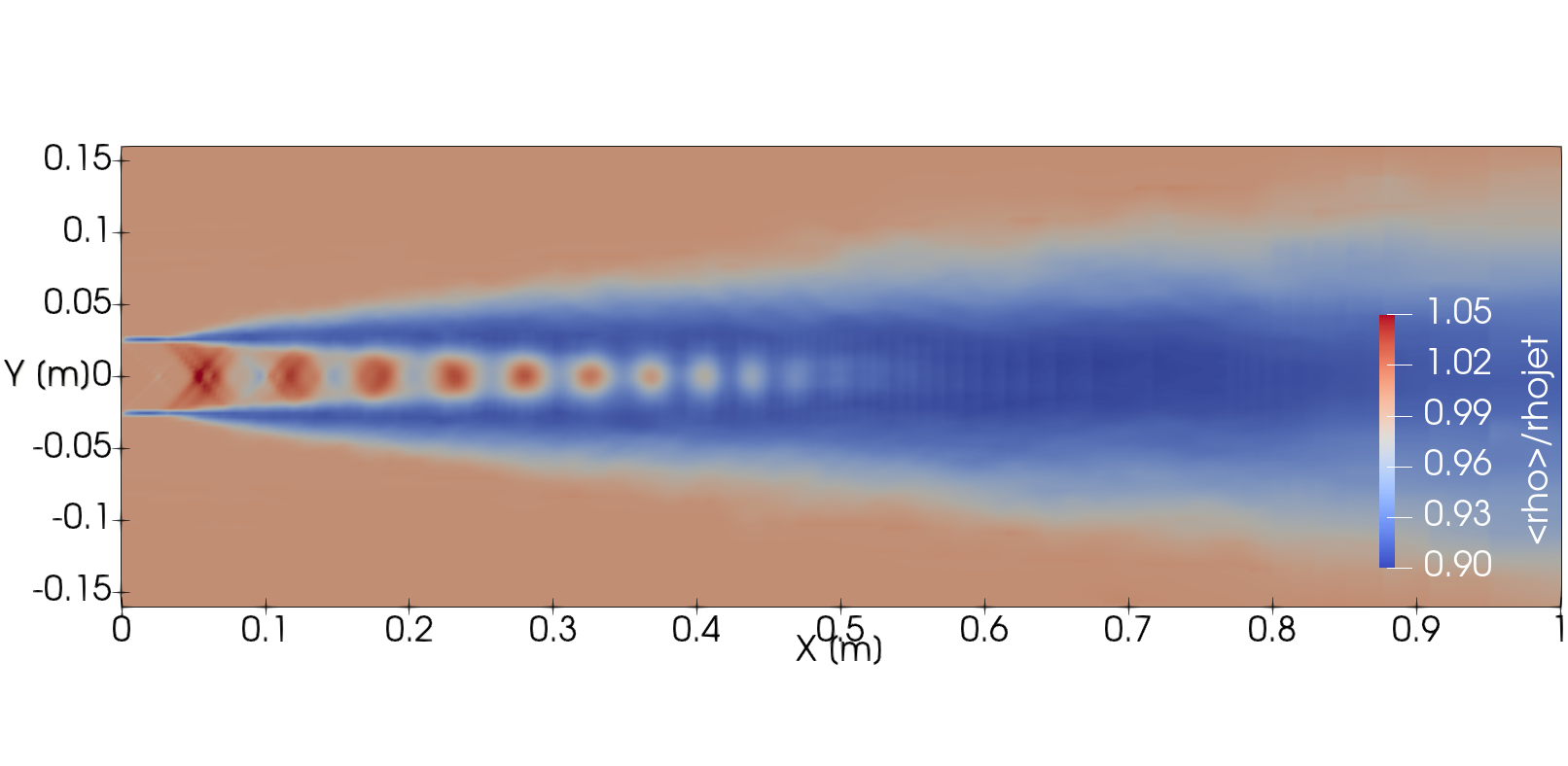}
	\label{fig.res_rhoc}
	}
\subfloat[S4 simulation.]{
	\includegraphics[trim = 0mm 35mm 0mm 50mm, clip, width=0.48\linewidth]{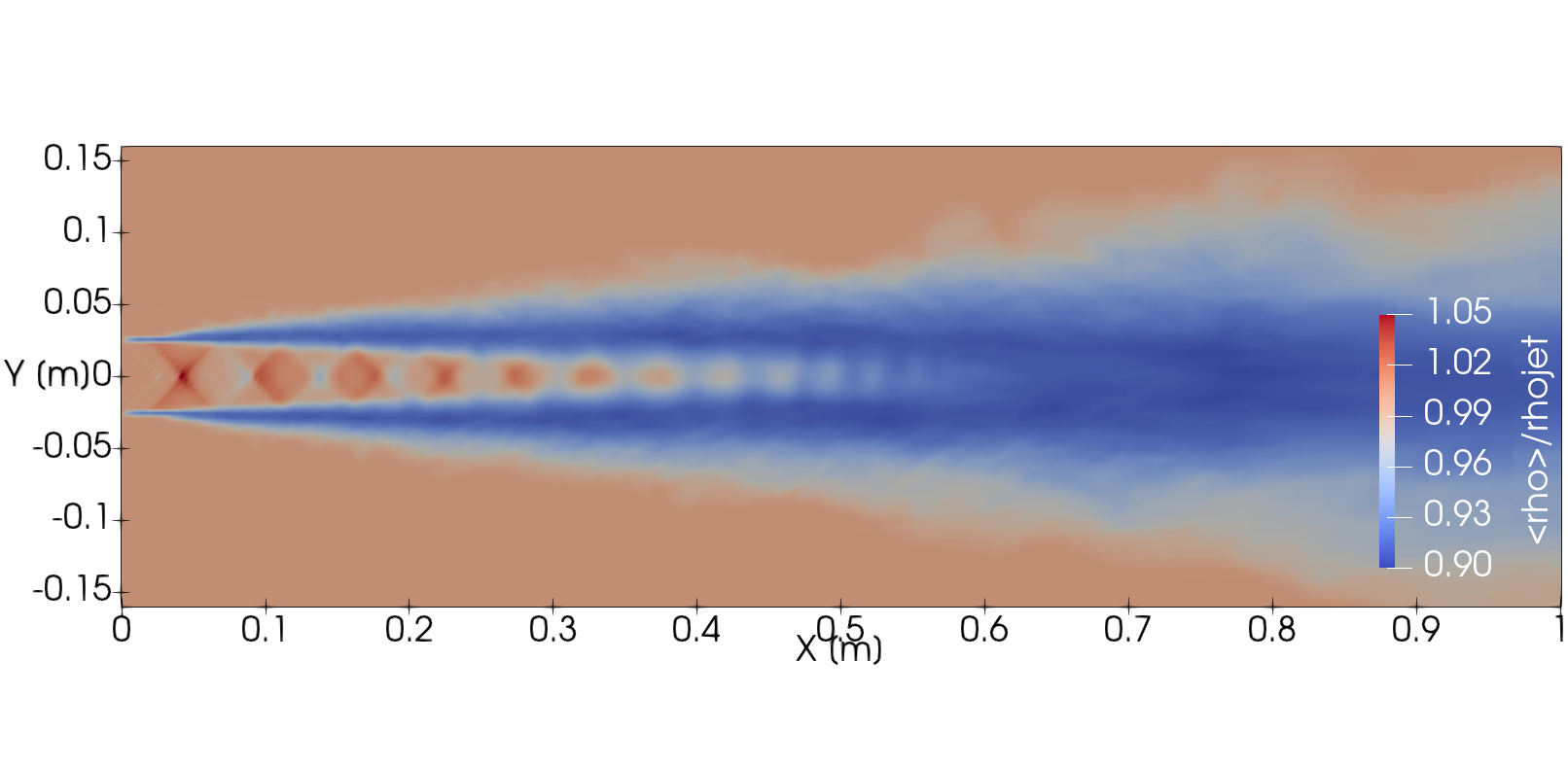}
	\label{fig.res_rhod}
	}
\caption{Contours of mean density on cutplane in $z/D=0$.}
\label{fig.res_rho}
\end{figure}

Finally, in Fig.\ \ref{fig.res_rho}, the contours of mean density are presented for all simulations. In these results, it is possible to better visualize the development of the series of shocks and expansion waves. Different from what has been observed in Figs.\ \ref{fig.res_mvelx} and \ref{fig.res_rvelx}, where the results from S1 and S2 simulation are very similar, in Figs.\ \ref{fig.res_rhoa} and \ref{fig.res_rhob} the results of mean pressure for S1 and S2 simulations, one can observe clear differences regarding the series of shock and expansion waves. In Fig.\ \ref{fig.res_rhoa} only three sets of shocks and expansion waves are clearly visible, while in Fig.\ \ref{fig.res_rhob} it is possible to observe more than 6 sets. It is possible to observe also that the sets of shocks and expansion waves from the S2 simulation are stronger than those from the S1 simulation. Analyzing Fig.\ \ref{fig.res_rhoc}, it is possible to observe that the S3 simulation produced even more sets of shocks and expansion waves than the S2 simulation, Fig.\ \ref{fig.res_rhob}, with larger intensity, that is evaluated by the level of variation of density produced by the shocks and expansion waves. Another aspect that can be observed is that the first set of shock and expansion waves in the S3 simulation is occurring closer to the jet inlet section than in S2 and S1 simulations, and appears to be a relation to the first set of shock and expansion waves with the beginning of the development of the shear layer. In Fig.\ \ref{fig.res_rhod} the results from S4 simulation is presented. One can observe the largest set of shock and expansion waves among all the simulations and also a thinner representation of the shocks and expansion waves, which can be closely related to the increased resolution of the simulation. It is also possible to observe a reduction in the intensity of the sets of shock and expansion waves when compared to the S3 simulation, Fig.\ \ref{fig.res_rhoc}. The first part of this section is concluded with the comparison of the contours of mean longitudinal velocity, RMS of longitudinal velocity fluctuation, and mean density among the simulations. In the second part, the numerical results are compared with experimental data.

\begin{figure}[htb!]
\centering
\subfloat[Centerline]{
	\includegraphics[width=0.47\linewidth]{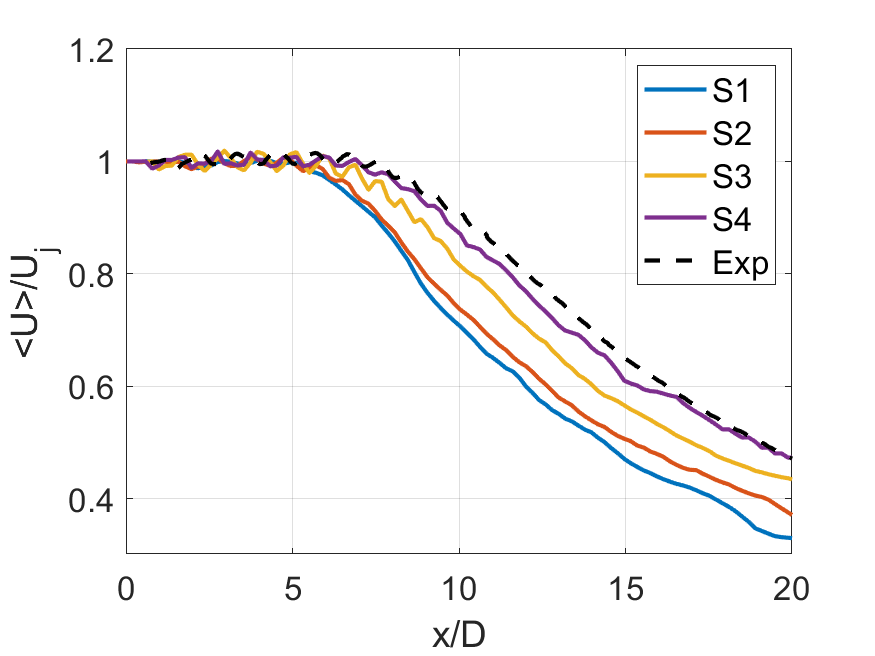}
	\label{fig.res1}	
	}%
\subfloat[Centerline]{
	\includegraphics[width=0.47\linewidth]{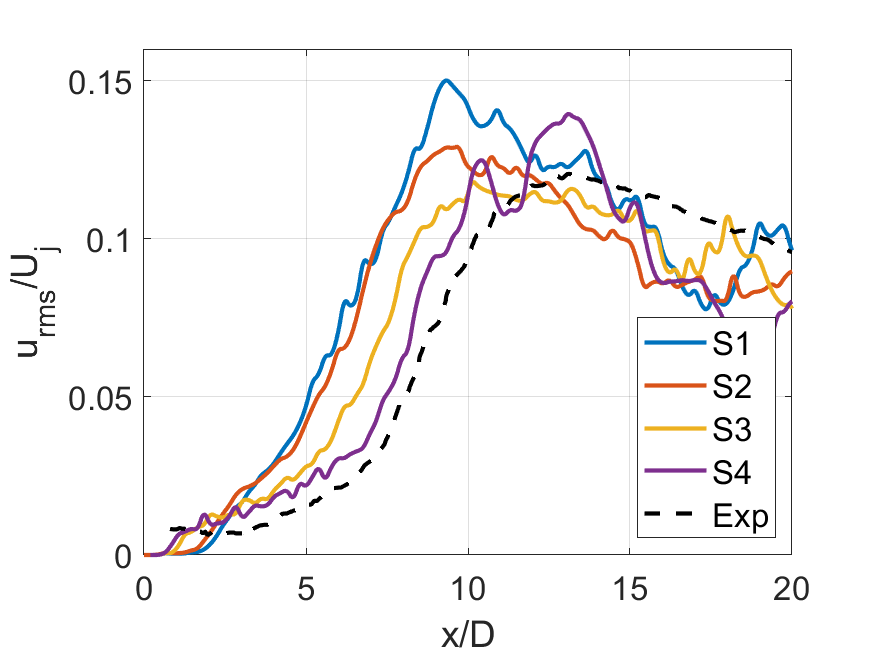}
	\label{fig.res2}	
	}
\newline
\subfloat[Lipline]{
	\includegraphics[width=0.47\linewidth]{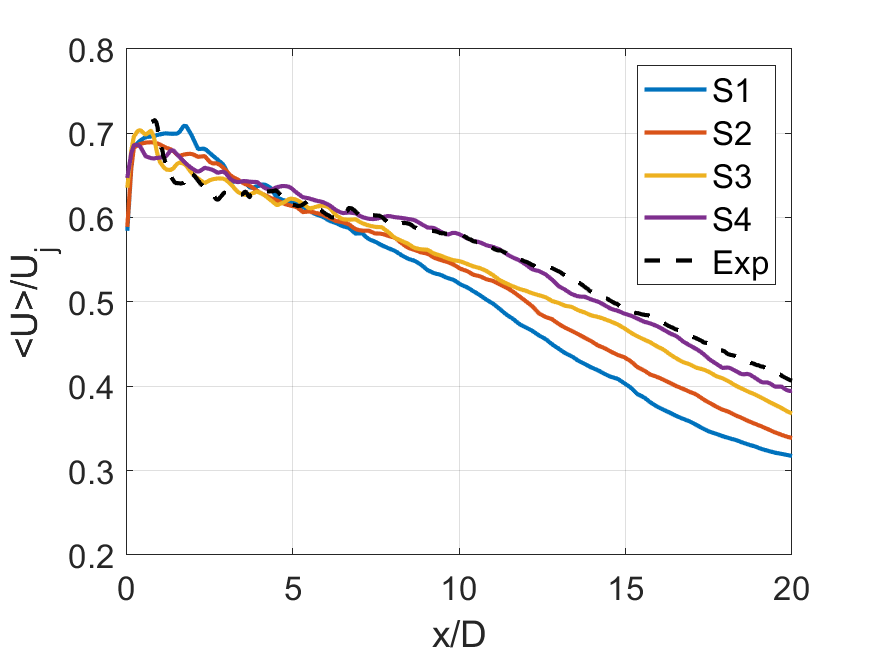}
	\label{fig.res3}
	}%
\subfloat[Lipline]{
	\includegraphics[width=0.47\linewidth]{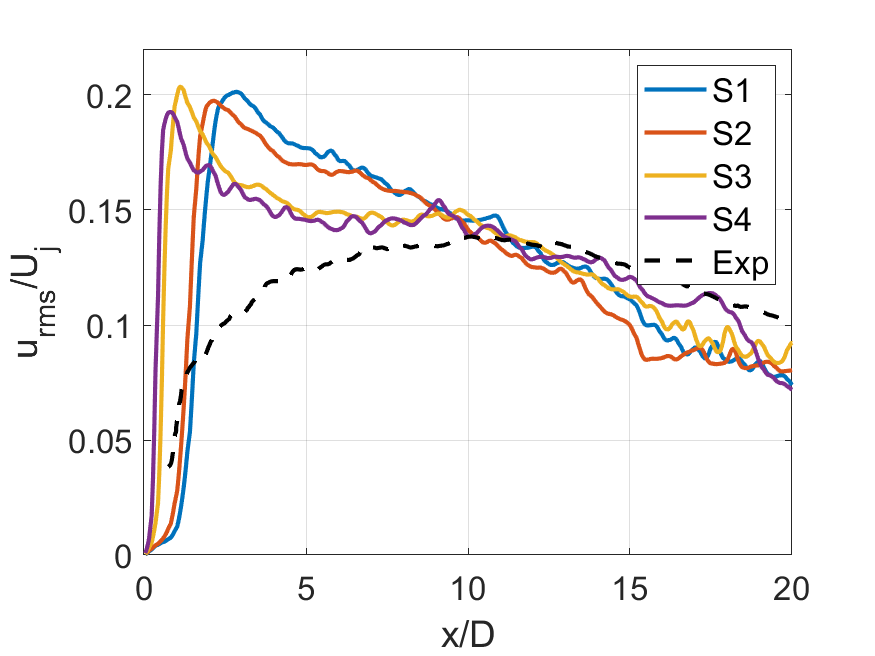}
	\label{fig.res4}	
	}
\caption{Results of mean streamwise velocity component distribution (left) and RMS of streamwise velocity fluctuation (right) in the jet centerline $y/D=0$ (top) and lipline $y/D=0.5$ (bottom).}
\label{fig.res1_4}
\end{figure}

In the second part, in Fig.\ \ref{fig.res1_4}, the numerical results of all simulations are compared to the experimental data. In Fig.\ \ref{fig.res1} the distribution of mean longitudinal velocity $<U>/U_j$ is presented in the centerline of the jet. One can observe in the figure that results from S1 and S2 simulations are almost equal. Results from the S3 simulation present a significant improvement when compared to previously performed simulations and the S4 simulation could almost capture the behavior observed in the experiments. In Fig.\ \ref{fig.res2} the RMS of longitudinal velocity fluctuation $u_{rms}/U_j$ is presented in the centerline of the jet. It is possible to observe once more in these results how well the resolution influenced the numerical results regarding the proximity to experimental data. The differences between S1 and S2 simulations are small, the S3 simulation got closer to experimental data and the S4 simulation once more presented the best match with experimental data. It can be observed in the results from the S4 simulation a double peak that does not appear in any other simulation or the experimental data. The authors believe that this result is a consequence of the fewer FFTs of the simulation in which data was gathered. Once more data could be used, authors believe that this characteristic would disappear and only one peak would be formed.

While the results for the jet centerline present always improvements in the simulations with increased resolution, in the lipline that behavior is not always observed. In Fig.\ \ref{fig.res3}, where the mean longitudinal velocity $<U>/U_j$ is presented in the lipline of the jet one may observe that far from the jet inlet section, the increased resolution produced a monotonic improvement in the numerical results, close to the jet inlet section, S3 simulation was the one that could better capture experimental data. However, it is not from the mean results where the greatest differences are observed. When analyzing the distribution of the RMS of longitudinal velocity distribution along the lipline of the jet, Fig.\ \ref{fig.res4}, one can observe that monotonically the increased resolution pushed the results away from experimental data. While in the experimental data it is possible to observe a smooth growth of RMS of longitudinal velocity fluctuation and almost a plateau from $x/D=5$ to $x/D=15$, in all the simulations there is a sudden increase in the RMS of longitudinal velocity fluctuation and after the peak, it is possible only to observe a reduction on the values. The authors believe that the differences observed in these results are related to the choice of the boundary condition imposed for the jet inlet that represents neither the boundary layer profile from the nozzle nor the turbulent intensity in the nozzle exit section. 

\begin{figure}[htb!]
\centering
\subfloat[$x/D=2.5$]{
	\includegraphics[trim = 30mm 0mm 30mm 0mm, clip, width=0.2\linewidth]{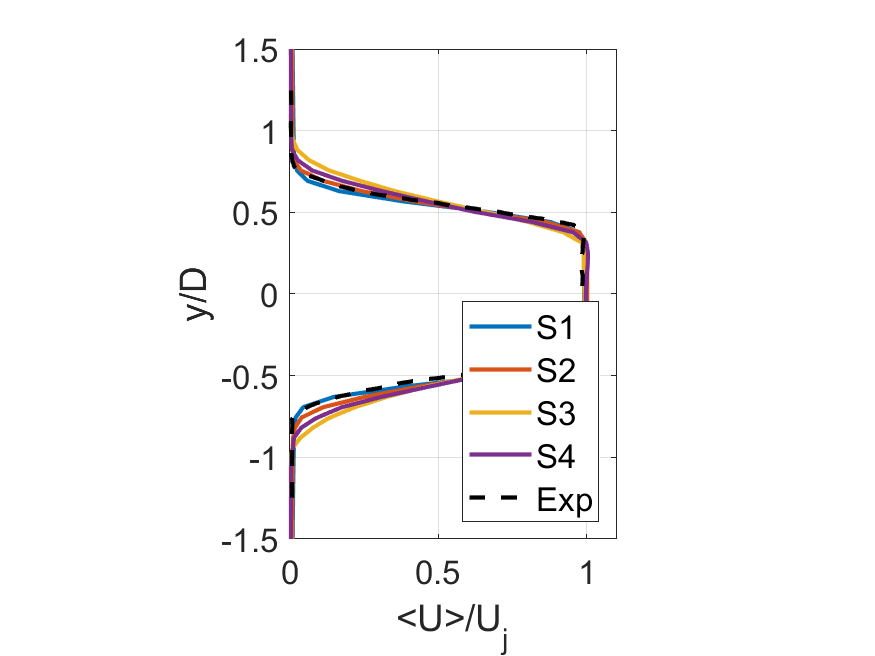}
	\label{fig.res5a}	
	}
\subfloat[$x/D=5$]{
	\includegraphics[trim = 30mm 0mm 30mm 0mm, clip, width=0.2\linewidth]{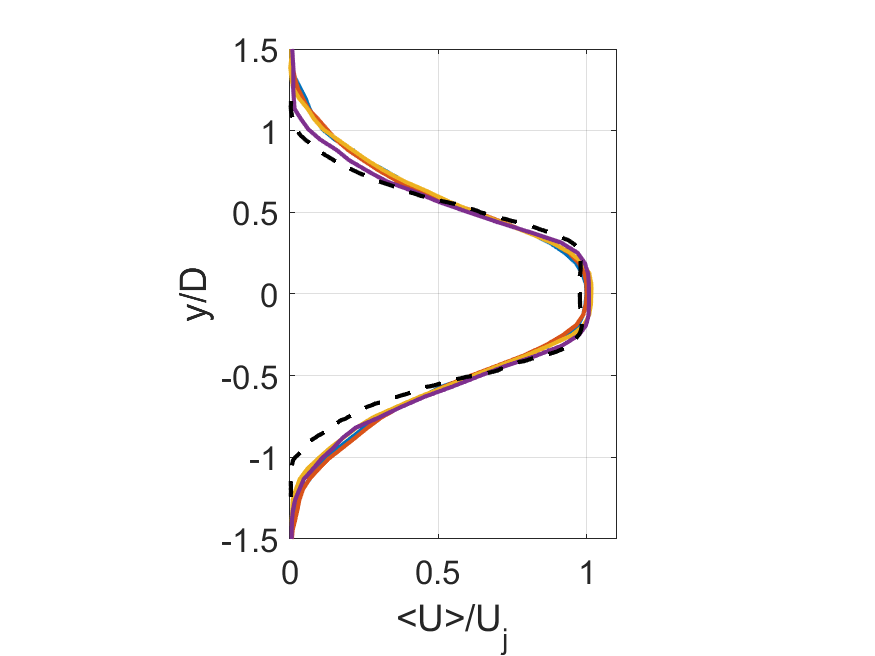}
	\label{fig.res5b}	
	}
\subfloat[$x/D=10$]{
	\includegraphics[trim = 30mm 0mm 30mm 0mm, clip, width=0.2\linewidth]{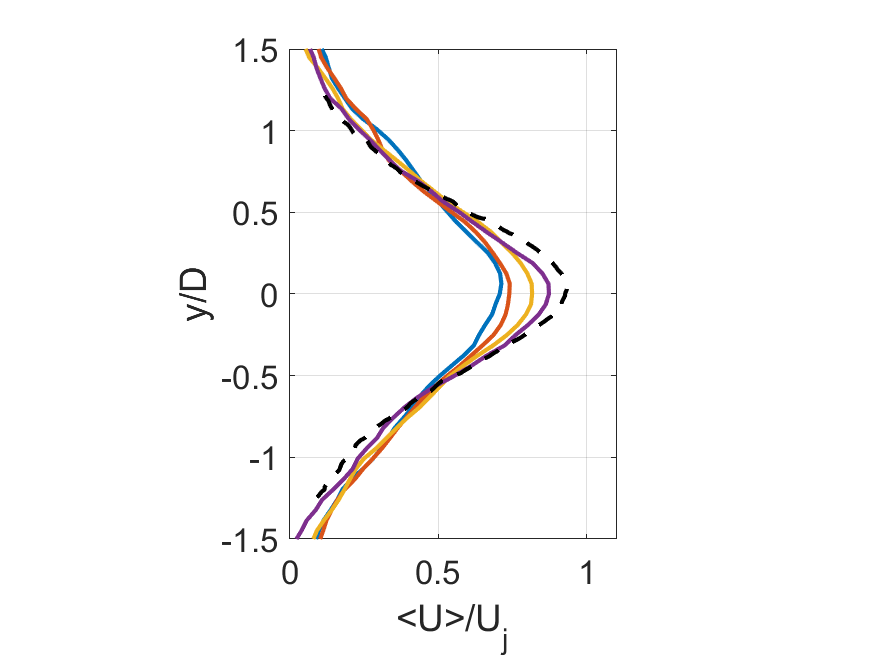}
	\label{fig.res5c}
	}
\subfloat[$x/D=15$]{
	\includegraphics[trim = 30mm 0mm 30mm 0mm, clip, width=0.2\linewidth]{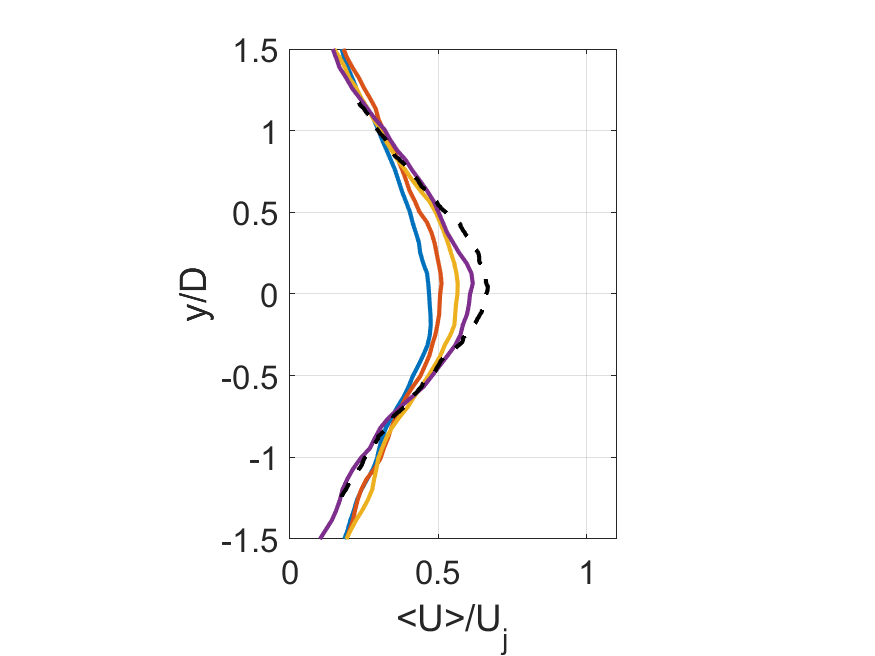}
	\label{fig.res5d}	
	}
\newline
\subfloat[$x/D=2.5$]{
	\includegraphics[trim = 30mm 0mm 30mm 0mm, clip, width=0.2\linewidth]{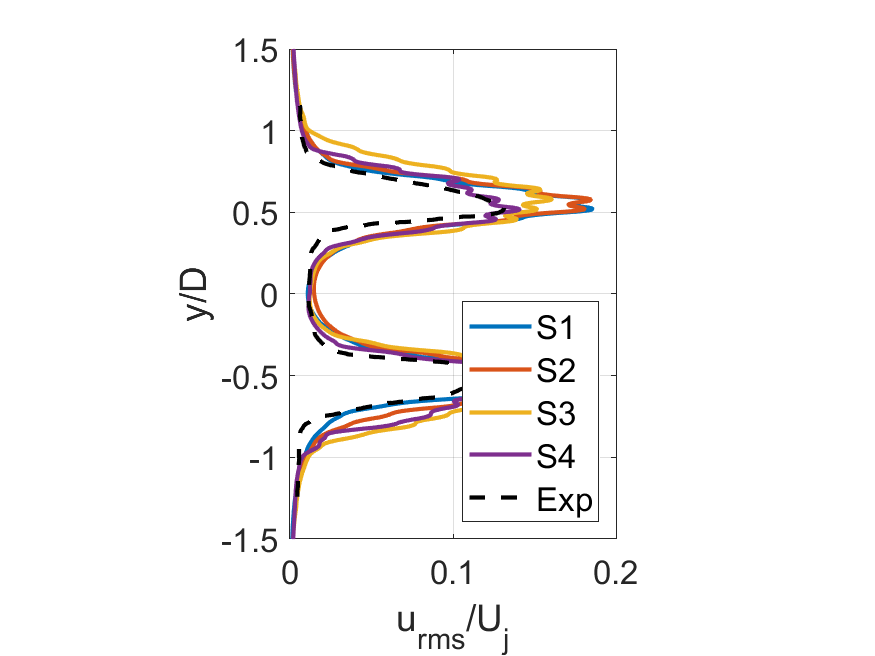}
	\label{fig.res5e}	
	}
\subfloat[$x/D=5$]{
	\includegraphics[trim = 30mm 0mm 30mm 0mm, clip, width=0.2\linewidth]{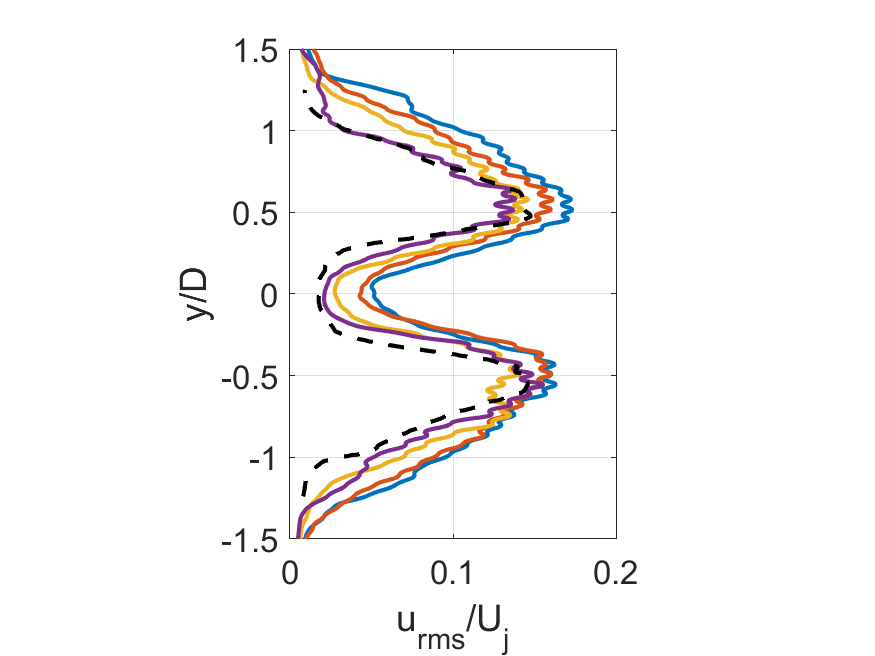}
	\label{fig.res5f}	
	}
\subfloat[$x/D=10$]{
	\includegraphics[trim = 30mm 0mm 30mm 0mm, clip, width=0.2\linewidth]{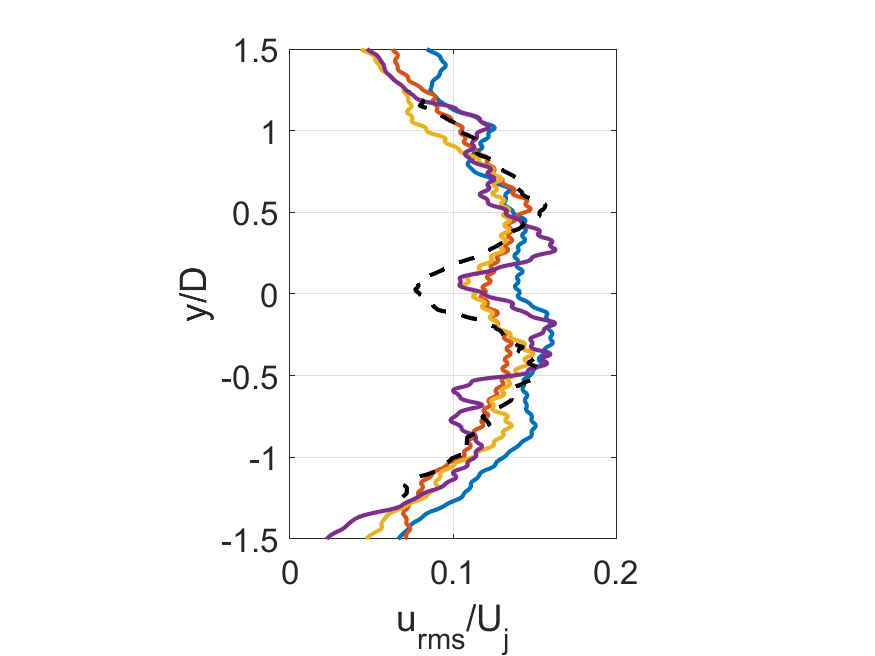}
	\label{fig.res5g}
	}
\subfloat[$x/D=15$]{
	\includegraphics[trim = 30mm 0mm 30mm 0mm, clip, width=0.2\linewidth]{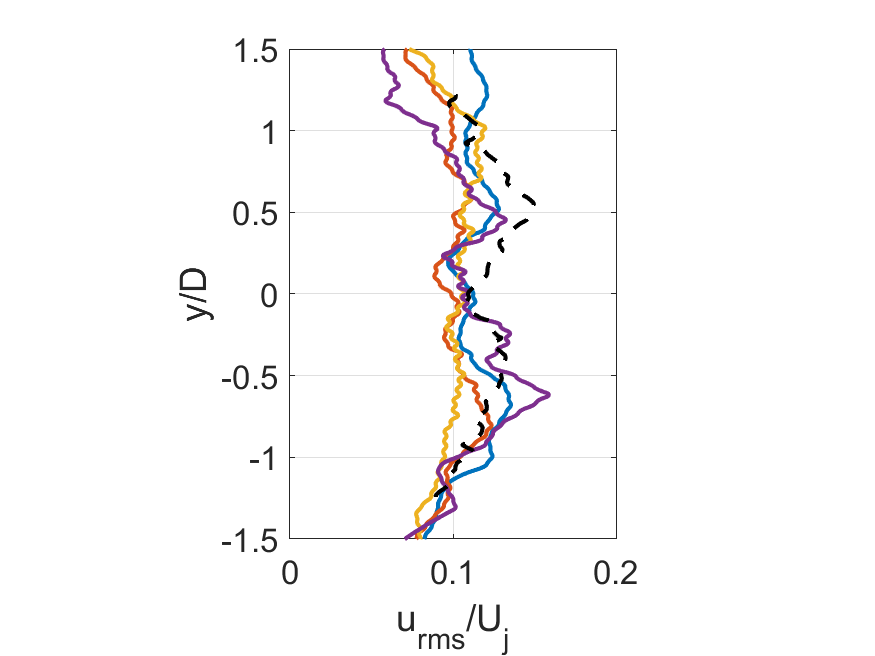}
	\label{fig.res5h}	
	}
\newline
\subfloat[$x/D=2.5$]{
	\includegraphics[trim = 30mm 0mm 30mm 0mm, clip, width=0.2\linewidth]{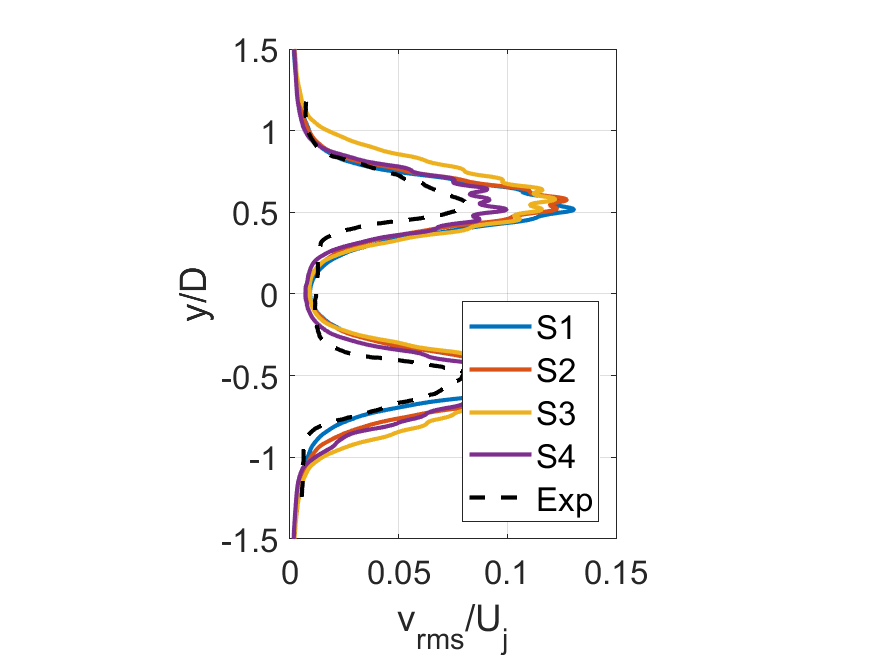}
	\label{fig.res5i}	
	}
\subfloat[$x/D=5$]{
	\includegraphics[trim = 30mm 0mm 30mm 0mm, clip, width=0.2\linewidth]{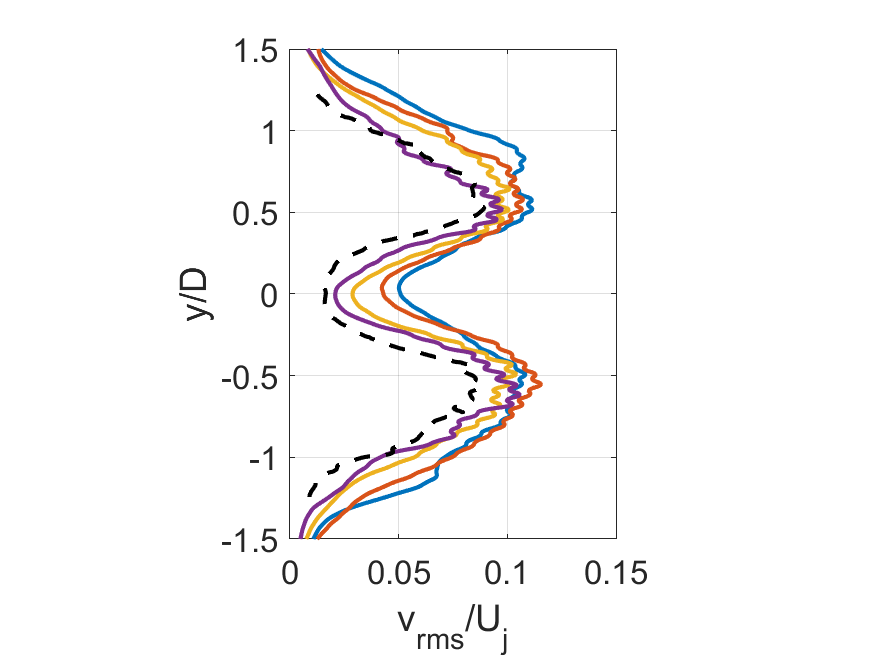}
	\label{fig.res5j}	
	}
\subfloat[$x/D=10$]{
	\includegraphics[trim = 30mm 0mm 30mm 0mm, clip, width=0.2\linewidth]{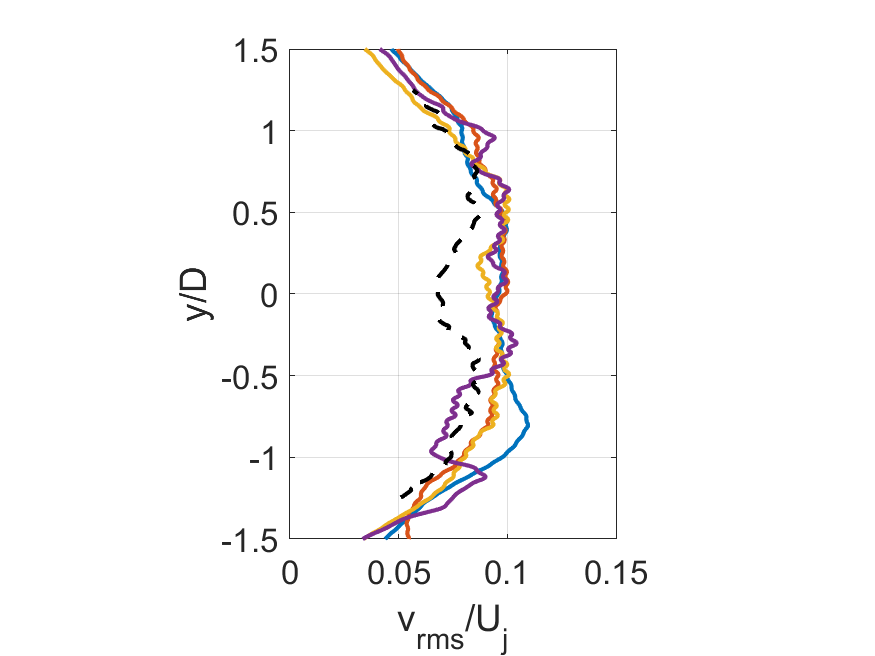}
	\label{fig.res5k}
	}
\subfloat[$x/D=15$]{
	\includegraphics[trim = 30mm 0mm 30mm 0mm, clip, width=0.2\linewidth]{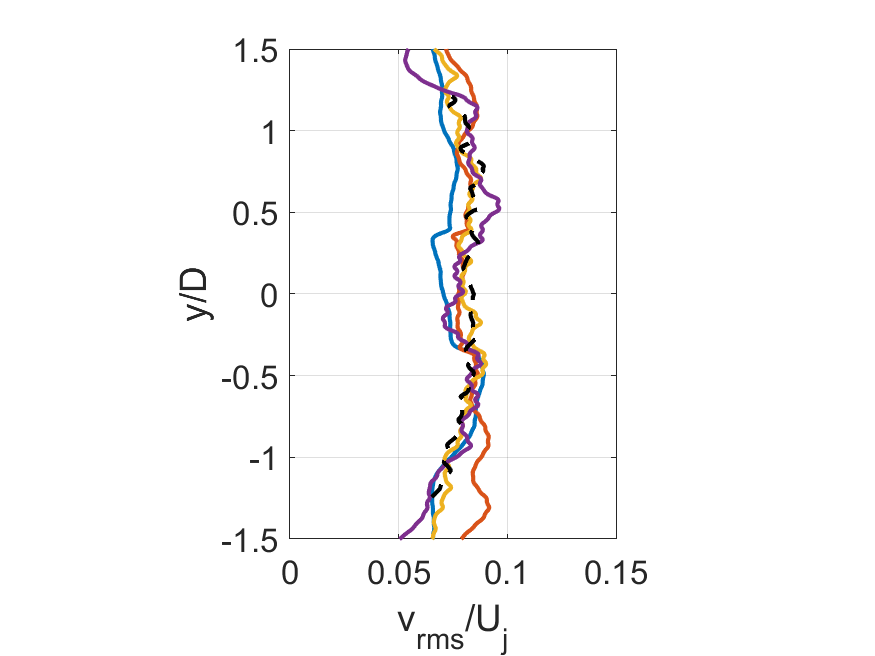}
	\label{fig.res5l}	
	}
\newline
\subfloat[$x/D=2.5$]{
	\includegraphics[trim = 30mm 0mm 30mm 0mm, clip, width=0.2\linewidth]{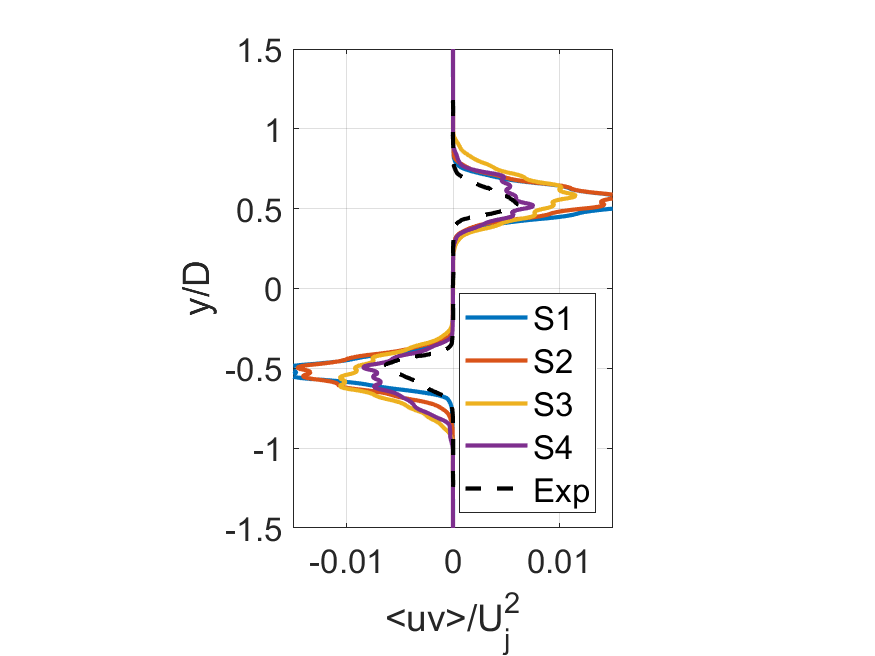}
	\label{fig.res5m}	
	}
\subfloat[$x/D=5$]{
	\includegraphics[trim = 30mm 0mm 30mm 0mm, clip, width=0.2\linewidth]{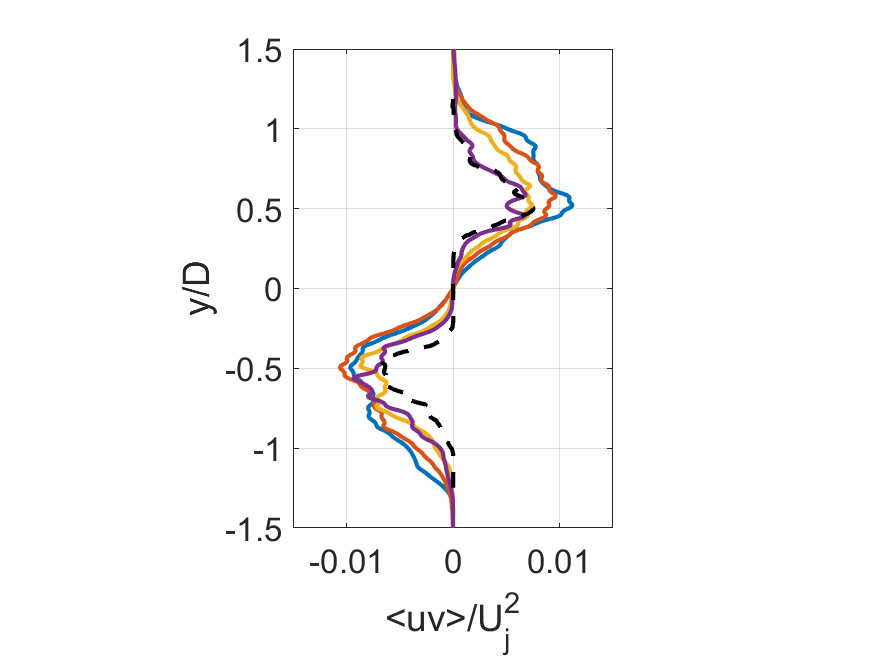}
	\label{fig.res5n}	
	}
\subfloat[$x/D=10$]{
	\includegraphics[trim = 30mm 0mm 30mm 0mm, clip, width=0.2\linewidth]{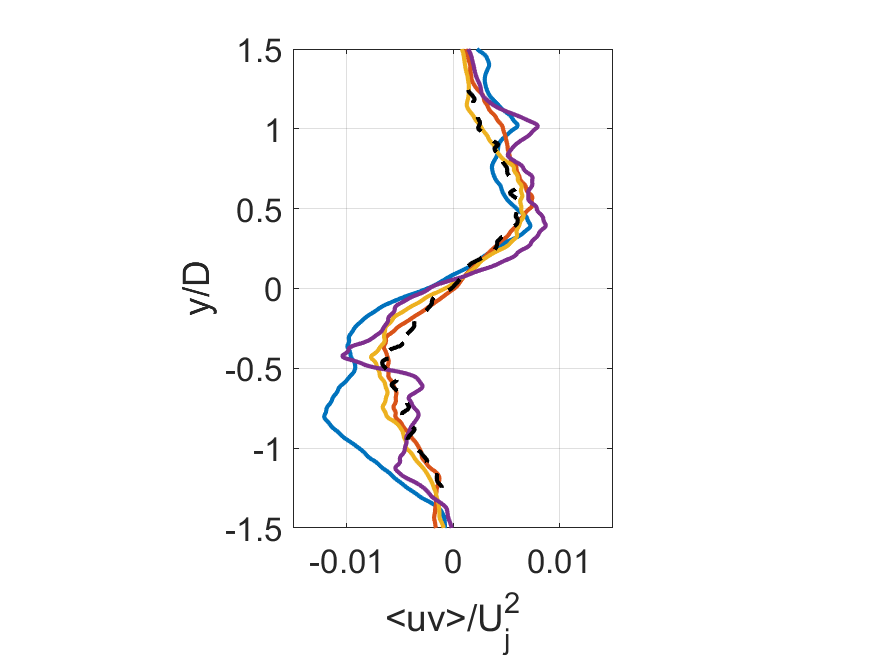}
	\label{fig.res5o}
	}
\subfloat[$x/D=15$]{
	\includegraphics[trim = 30mm 0mm 30mm 0mm, clip, width=0.2\linewidth]{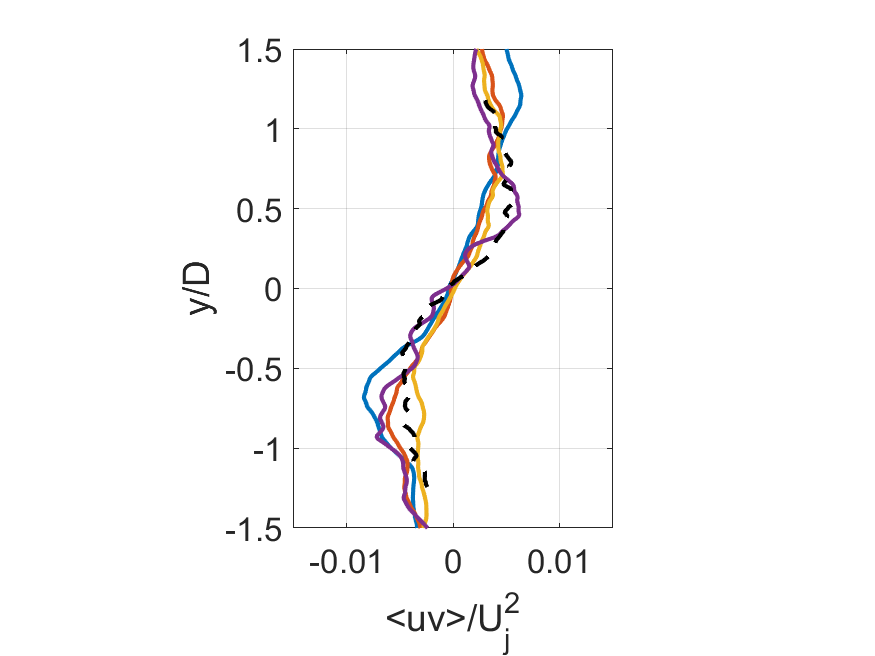}
	\label{fig.res5p}	
	}
\newline
\caption{Profiles of mean streamwise velocity component, RMS of streamwise velocity 
fluctuation, RMS of radial velocity fluctuation, and mean shear-stress tensor 
component (from top to bottom) at four streamwise positions $x/D=2.5$, $x/D=5$, $x/D=10$ and 
$x/D=15$ (from left to right).}
\label{fig.res5}
\end{figure}

The results in Fig.\ \ref{fig.res5} present different statistical properties of the flow in different longitudinal positions. The first set of results, in Figs.\ \ref{fig.res5a} to \ref{fig.res5d}, concerns the mean of longitudinal velocity. The S1 simulation is in agreement with the experimental data at $x/D=2.5$, Fig.\ \ref{fig.res5a}. In the position $x/D=5$, Fig.\ \ref{fig.res5b} all the simulations produce very similar results. Moving forward and analyzing the results in position $x/D=10$, Fig.\ \ref{fig.res5c} it is possible to observe a similar behavior between S1 and S2 simulations, S3 simulation presenting improvements to the other two and S4 simulation presenting the best match with experimental data. In the last position $x/D=15$, Fig.\ \ref{fig.res5a}, a monotonically improvement is observed with increased resolution the the simulations.

\FloatBarrier

The profiles of RMS values of longitudinal velocity fluctuation are presented in Figs.\ \ref{fig.res5e} to \ref{fig.res5h}. The simulation results at $x/D=2.5$, Fig.\ \ref{fig.res5e}, present a similar profile with differences only in the peak of RMS of longitudinal velocity fluctuation with a monotonically decrease in the peak with increased resolution and proximity to experimental data. Similar behavior is observed in the next position at $x/D=5$, Fig.\ \ref{fig.res5f}. In the position $x/D=10$, Fig.\ \ref{fig.res5g} the main aspects of the flow are captured except for the smaller values of RMS of longitudinal velocity fluctuation in the center of the jet that is only badly captured by S4 simulation. The results for all simulations and experimental data are very similar in the last position $x/D=15$, Fig.\ \ref{fig.res5h}

Profiles of RMS values of radial velocity component fluctuation are presented in Figs.\ \ref{fig.res5i} to \ref{fig.res5l}. They exhibit similar behavior as the longitudinal velocity fluctuation. One can also observe the positive effects of the increased resolution on the profiles of the mean shear-stress tensor component, Figs.\ \ref{fig.res5m} to \ref{fig.res5p}. The profiles from the S4 simulation are in good agreement with the experimental data and they indicate considerable improvement when compared to the simulations with smaller resolution.

This result concludes our analysis of the numerical results from the simulations. It was possible to observe, in general, that the improved resolution of the simulations produced better results compared to experimental data. The simulation with the highest resolution, the S4 simulation, was the one that better matched the experimental data. Only in the jet lipline, this behavior is not observed. The authors have strong confidence that this has nothing to do with the effect of the resolution, instead, it may be strongly related to the choice of the boundary condition for the jet inlet condition, that does not reproduce the boundary layer developed inside the nozzle nor the turbulent intensity in the region. To improve the quality of the simulations in the jet lipline, a new boundary condition or a different simulation strategy should be adopted.

\subsection{ANALYSIS OF COMPUTATIONAL EFFORT}
At this point of the work, it is important to discuss some other aspects of computational effort to be able to improve the computational efficiency of the simulations. The main parameter utilized to measure the efficiency of a simulation is the Performance Index $PID$, which can be calculated by
\begin{equation}
PID = \frac{wall \hspace{2pt} clock \hspace{2pt} time \hspace{3pt} n_{cores}}{n_{DOF} \hspace{3pt} n_{time \hspace{1pt} steps} \hspace{3pt} n_{RK-stages}},
\end{equation}
where $wall \hspace{2pt} clock \hspace{2pt} time$ is the time the simulation needed to perform $n_{time \hspace{1pt} steps}$ time steps, $n_{cores}$ is the quantity of cores used in the simulation, $n_{DOF}$ is the number of DOF of the simulation and $n_{RK-stages}$ is the number of stages from the Runge-Kutta scheme. The PID was calculated for all four simulations and the results are presented in Tab.\ \ref{tab.pid}.

\begin{table}[htb!]
\centering
\caption{Summary of Performance Index $PID$ from all simulations.}
\begin{tabular}{ c | c }
Simulation & PID ($\mu s$) \\ \hline
S1 & $8$ \\
S2 & $15$ \\
S3 & $5$ \\
S4 & $2$ \\ \hline
\end{tabular}
\label{tab.pid}
\end{table}

It is important to clarify to the author that the numerical solver presented some improvements during the execution of the simulations and they can be related to the improvements in the $PID$ reduction from S3 and S4 simulations compared to S1 and S2 simulations. It is also possible to argue that with the increased number of degrees of freedom it is expected that more computation is performed with a similar number of cores, which also should increase the efficiency of the simulations. If we compare only S1 and S2 simulations, for the same number of DOF, the third-order accurate simulations cost almost twice the effort of a second-order accurate simulation. If these values are employed in the two simulations procedure performed in this analysis, the total cost of the simulation can be compared.

The first simulation procedure involves the whole calculation of the 9 FTT with a mesh of $50 \times 10^6$ elements that produces $\approx 400 \times 10^6$ DOF when simulated with second-order accurate discretization, which could produce a very similar result to those of S4 simulation. If the procedure for S4 simulation could be completely performed, it could initially start its 5 FTT with second-order accurate discretization with a total of $120 \times 10^6$ DOF. At this point, we do not consider the effect of the number of degrees of freedom in the PID, only the order of discretization. For this 5 FTT, the total time would be $3.34 \times$ smaller than those to perform the first simulation procedure due to the reduced number of degrees of freedom. Then, it is possible to consider the next 4 FTT calculated with third-order accurate discretization and a total of $ \approx 400 \times 10^6$ DOF. In this 4FTT the cost of the third-order accurate discretization is twice of the second-order accurate simulation. If the total time of the second procedure is calculated, it costs $\approx 5 \%$ more than the procedure with second-order accuracy.

What is discussed here is that, once it is possible to start the high-order simulation with a previous result from another order of accuracy with the same mesh, it is possible to reduce the time required to obtain the desired data with high-order simulation and consequently reduce the cost of the total simulation. In the proposed procedure, the cost of the total third-order accurate simulation was only $5 \%$ larger than those of a second-order accurate simulation. This result is expressive and very interesting for high-order simulations.

Another important point to present is that the computational code utilized presented very good scalability with the number of cores. The tests for the S4 simulation varied the number of cores from a few hundred cores to a few thousand cores and the $PID$ was always close to $2 \mu s$. 

\FloatBarrier

\section{CONCLUDING REMARKS}
In this work, the employment of a discontinuous Galerkin framework called FLEXI was investigated for the LES simulation of supersonic free round jets. A total of four simulations are performed with 3 different meshes and second and third-order accuracy. The range of degrees of freedom from the simulations varies from $50 \times 10^6$ to $400 \times 10^6$. All the simulations are performed for the same geometric model and with the same boundary conditions.

The results of the simulations are firstly compared visually between themselves to compare how they are capturing the main features of the flow: extension of the high-velocity region, development of the shear layer, and development of the sets of shocks and expansion waves. The results showed that with increased resolution the high-velocity regions got longer. The development of the shear layer starts closer to the jet inlet section and presents a smaller spreading. The number of sets of shocks and expansion waves increased and the visual of the shocks and expansion waves are thinner.

The numerical results showed that, in general, the increase in the resolution of the simulation, especially the number of degrees of freedom, produced better results when compared to experimental data. This behavior is observed in the results of mean longitudinal velocity distribution and RMS of longitudinal velocity fluctuation distribution in the centerline. It is also observed in the results from the four spanwise planes. The only results that do not follow this behavior are the mean longitudinal and RMS of longitudinal velocity fluctuation distributions along the jet lipline. In these regions, with increased resolutions, the results are pushed away from experimental data. The authors believe this behavior is related to the lack of resolution from the jet inlet boundary condition that could not reproduce the boundary layer and turbulent intensity from the experiments.

The analysis of the computational effort of the simulation showed that even utilizing a high-order method that costs more than a second-order method for the same number of degrees of freedom it was possible to reproduce a third-order simulation with only $5 \%$ more computational cost of total simulation by initializing the simulation with a smaller order of accuracy.

The work reached is objective of identifying the guidelines for performing LES simulations of supersonic jet flows using a discontinuous Galerkin scheme with adequate results with a reasonable computational cost. The open point on the jet inlet condition is the next step in the development of the work. 

\section*{ACKNOWLEDGMENTS}
The authors acknowledge the support for the present research provided by Conselho Nacional de Desenvolvimento Cient\'{\i}fico e Tecnol\'{o}gico, CNPq, under the Research Grant No.\ 309985/2013-7\@. The work is also supported by the computational resources from the Center for Mathematical Sciences Applied to Industry, CeMEAI, funded by Funda\c{c}\~{a}o de Amparo \`{a} Pesquisa do Estado de S\~{a}o Paulo, FAPESP, under the Research Grant No.\ 2013/07375-0\@. The authors further acknowledge the National Laboratory for Scientific Computing (LNCC/MCTI, Brazil) for providing HPC resources of the SDumont supercomputer. This work was also granted access to the HPC resources of IDRIS under the allocation 2020-A0092A12067 / 2021-A0112A12067 made by GENCI. The first author acknowledges authorization by his employer, Embraer S.A., which has allowed his participation in the present research effort. The doctoral scholarship provide by FAPESP to the third author, under the Grant No.\ 2018/05524-1\@, is thankfully acknowledged. Additional support to the fourth author under the FAPESP Research Grant No.\ 2013/07375-0\@ is also gratefully acknowledged. This study was financed in part by the Coordenação de Aperfeiçoamento de Pessoal de Nível Superior - Brasil (CAPES) - Finance Code 001.

\bibliographystyle{plain}
\bibliography{bibfile_paper}

\end{document}